\documentclass[graybox]{svmult}

\usepackage{bbm}
\usepackage{booktabs}
\usepackage[labelformat=parens]{subfig}
\usepackage{type1cm}        
\usepackage{makeidx}         
\usepackage{graphicx}        
\usepackage{multicol}        
\usepackage[bottom]{footmisc}

\usepackage[Export]{adjustbox}
\usepackage{newtxtext}       %
\usepackage{newtxmath}       

\usepackage{natbib}

\setcitestyle{numeric}
\makeindex             

\begin{document}

\title*{Sketch-based Creativity Support Tools using Deep Learning}
\author{Forrest Huang, Eldon Schoop, David Ha, Jeffrey Nichols, John Canny}
\institute{Forrest Huang \at University of California, Berkeley
\email{forrest\_huang@berkeley.edu}
\and Eldon Schoop \at University of California, Berkeley
\email{eschoop@berkeley.edu}
\and David Ha \at Google Brain
\email{hadavid@google.com}
\and Jeffrey Nichols \at Apple, Inc. \email{jeff@jeffreynichols.com}
\and John Canny \at University of California, Berkeley
\email{canny@berkeley.edu}}
%
%
\newcommand{\CN}{{\color{blue}[Citation Needed]}}

\newcommand\sig[1]{#1}
\newcommand\crd[1]{#1}
\newcommand{\systemname}{Scones~}
\newcommand{\systemnamenospace}{Scones}

\maketitle
\vspace{-4em}
This is a preprint of the following chapter: Huang et al., Sketch-based Creativity Support Tools using Deep Learning, published in Artificial Intelligence for Human Computer Interaction: A Modern Approach, edited by Yang Li and Otmar Hilliges, 2021, Springer reproduced with permission of Springer Nature Switzerland AG. The final authenticated version is available online at:\\ https://doi.org/10.1007/978-3-030-82681-9\\
\\

\abstract*{Sketching is a natural and effective visual communication medium commonly used in creative processes. Recent developments in deep-learning models drastically improved machines' ability in understanding and generating visual content. An exciting area of development explores deep-learning approaches used to model human sketches, opening opportunities for creative applications. This chapter describes three fundamental aspects of researches that develop deep-learning-driven creativity support tools that consume and generates sketches: 1) a data collection effort that generated a new paired dataset between sketches and mobile user interfaces; 2) a sketch-based user interface retrieval system adapted from state-of-the-art computer vision techniques; and, 3) a . We present the research Each chapter should be preceded by an abstract (no more than 200 words) that summarizes the content. The abstract will appear \textit{online} at \url{www.SpringerLink.com} and be available with unrestricted access. This allows unregistered users to read the abstract as a teaser for the complete chapter.
Please use the 'starred' version of the \texttt{abstract} command for typesetting the text of the online abstracts (cf. source file of this chapter template \texttt{abstract}) and include them with the source files of your manuscript. Use the plain \texttt{abstract} command if the abstract is also to appear in the printed version of the book.}

\abstract{Sketching is a natural and effective visual communication medium commonly used in creative processes. Recent developments in deep-learning models drastically improved machines' ability in understanding and generating visual content. An exciting area of development explores deep-learning approaches used to model human sketches, opening opportunities for creative applications. This chapter describes three fundamental steps in developing deep-learning-driven creativity support tools that consumes and generates sketches: 1) a data collection effort that generated a new paired dataset between sketches and mobile user interfaces; 2) a sketch-based user interface retrieval system adapted from state-of-the-art computer vision techniques; and, 3) a conversational sketching system that supports the novel interaction of a natural-language-based sketch/critique authoring process. In this chapter, we survey relevant prior work in both the deep-learning and human-computer-interaction communities, document the data collection process and the systems' architectures in detail, present qualitative and quantitative results, and paint the landscape of several future research directions in this exciting area.}

\newpage

\section{Introduction}
\label{sec:intro}


Sketching is a natural and effective 
means to express
novel artistic and functional concepts. It is an integral part of the
creative process for many artists, engineers, and educators.
The abstract yet expressive nature of sketches enables sketchers to quickly communicate conceptual and high-level ideas visually 
while leaving out unnecessary details. These characteristics are most notably manifested in the use of sketches in design processes, where sketches are used by designers to iteratively discuss and critique high-level design concepts and ideas.

Recent advances in deep-learning (DL) models greatly improved machines' abilities to perform Computer Vision tasks. In particular, convolutional neural networks, recurrent neural networks, and attention mechanisms dramatically outperform prior state-of-the-art methods in comprehending and generating visual content.
They can perform these tasks even conditioned on user-specified natural language or other accompanying semantic information. These architectures provide great opportunities for developing creativity-support applications of the type we have argued for. They support sketch inputs and outputs for applications such as sketch-based image retrieval and sketch generation systems.

This chapter explores and surveys multiple facets of research in deep-learning-based sketching systems that support creative processes. We describe three research projects spanning design and artistic applications, targeting amateur and professional users, and using sketches as inputs and outputs. We outline three important aspects in this area of research: 1) collecting appropriate sketch-based data; 2) adapting existing architectures and tasks to the sketch domain; and, 3) formulating new tasks that supports novel interactions using state-of-the-art model architectures and deploying and evaluating these novel systems. 

The key projects that we will include in this chapter are as follows:
\begin{svgraybox}
\begin{enumerate}
    \item a new dataset consisting of paired sketches and UI designs produced by UI/UX designers
    \item a new system that retrieves relevant UI designs based on user-specified sketches that can support various design applications
    \item a novel formulation of an iterative sketch-authoring process based on user-specified natural language inputs
\end{enumerate}
\end{svgraybox}

We begin the chapter by reviewing relevant prior literature on sketch-based applications in both the Deep-Learning (DL) community and Human-Computer-Interaction (HCI) community. We then describe the three aforementioned projects in detail and present qualitative and quantitative experimental results on existing and novel tasks. Finally, we conclude the chapter with a discussion of several avenues of further research in sketch-based deep-learning systems. We hope to provide a guide to aid the research and development of newer systems that works with sketches and sketch-based tasks. We aim to push the frontier of future sketch-based systems research to allow machines to support sketch-based creativity in various domains for users with all levels of expertise.



\section{Role of Sketching in supporting Creative Activities}
\label{sec:role_sketching}

Sketching plays a pivotal role in many types of creative activities because of its highly visual nature and its flexibility for creation and manipulation: users can create and imagine any kind of visual content, and continuously revise it without being constrained by unnecessary details. Sketches are an independent art form, but are also extensively used to draft and guide other forms of artistic expression such as oil painting, or storyboarding in films and motion graphics. Moreover, because sketches effectively communicate visual ideas, they are well-suited for design processes such as User Interface (UI) or User Experience (UX) Design. 

For these reasons, a plethora of research systems and tools that use sketches have been developed by the HCI community. We survey several notable systems in the domains of artistic sketches and design sketches in this section.

\subsection{Sketch-based Applications supporting Artistic Expressions}
Prior works that aim to support artistic sketches have mostly taken the form of real-time assistance to directly improve the final sketching product, or to generate sketching tutorials for improving users' sketching proficiency. A number of sketching assistants use automatically-generated and crowd-sourced drawing guidance. ShadowDraw \citep{shadowdraw} and EZ-sketching \citep{ez-sketching} use edge images traced from natural images to suggest realistic sketch strokes to users. PortraitSketch provides sketching assistance specifically on facial sketches by adjusting geometry and stroke parameters~\citep{potrait-sketch}. Real-time, crowd-sourced feedback have also been used to to correct and improve users' sketched strokes~\citep{drawingCrowdsourcing}. 

In addition to assisted sketching tools, researchers have developed tutorial systems to improve users' sketching proficiency. How2Sketch automatically generates multi-step tutorials for sketching 3D objects~\citep{how2sketch}. Sketch-sketch revolution provides first-hand experiences created by sketch experts for novice sketchers~\citep{sketch-sketch} . 

\subsection{Sketch-based Applications supporting Design in Various Domains}

Designers use sketches to expand novel ideas, visualize abstract concepts, and rapidly compare alternatives~\citep{buxton}. They are commonplace in the modern design workspace and typically require minimal effort for designers to produce. They are also sometimes preferred over high-fidelity artifacts because the details left out by abstract sketches implies and indicates incompleteness of designs. They encourage designers to more freely imagine and provide alternatives based on current designs without being concerned about committed to existing designs. From some of our informal conversations with designers, it is observed that designers will trace high-fidelity design renderings using rough sketch strokes for soliciting higher-level, creative feedback.

Research in the \crd{HCI community} has produced interfaces that use drawing input for creating interactive design prototypes. SILK is the first system that allows designers to author interactive, low-fidelity UI prototypes by sketching~\citep{silk}. DENIM allows web designers to prototype with sketches at multiple detail levels~\citep{denim}. More recently, Apparition uniquely allow users sketch their desired interfaces while having crowdworkers translate sketches into workable prototypes in near real-time~\citep{apparition}.

\section{Large-scale Sketch Datasets and Sketch-based Deep-learning Applications}
With the widespread use of DL approaches in Computer Vision, researchers have collected large-scale sketch datasets to train and evaluate these models.
We describe several projects in the DL community that enable sketch-based retrieval and sketch generation tasks using recently-collected sketch datasets. The capabilities of these models allow us to create novel interactive systems that were difficult prior to the introduction of deep-learning.

\subsection{Large-scale Sketch Datasets}
\label{sec:datasets}
To support DL-based approaches towards sketch comprehension and generation, which rely heavily on large-scale datasets, researchers have crowdsourced sketch datasets of individual objects. Most of these datasets have focused on individual instances that either correspond to natural language or a general semantic class. The Quick, Draw!~\citep{quickdraw} and TU-Berlin~\citep{tuberlin-sketch} datasets consist of human-drawn sketches for 345 and 250 object classes respectively. SketchyDB provides paired images and simple sketches for retrieval tasks~\citep{sketchy}.

More recently, researchers have explored beyond individual sketched instances to compile datasets of sketched multi-object scenes. \crd{The SketchyScene dataset consists of sketched scenes of pre-drawn objects transformed and resized by humans, as scene sketches are highly demanding for users to create from scratch~\citep{sketchyscenes}.} SketchyCOCO starts with the MSCOCO dataset and retrieves relevant instances from the Sketchy dataset, compiling them into complex scenes~\citep{sketchycoco}. Nevertheless, none of these datasets consist of sketched scenes drawn by humans entirely from scratch. This is mainly due to the high skill barrier of generating these sketches that make large-scale crowd-sourcing difficult~\citep{sketchyscenes}. 

Related to complex sketches of multiple artistic objects, the DiDi dataset introduced drawings of flow-charts and functional diagrams traced entirely by human users~\citep{didi}. While users did not design the diagrams from scratch, they generated each individual sketch stroke based on the procedurally generated templates.

\subsection{Sketch-Based Image and 3D Model Retrieval}
\label{sec:sbir}
Sketch-based Image Retrieval is a frequently studied problem in the Computer Vision community. The current state-of-the-art solution to an instance of this problem is deep-learning-based~\citep{sbirsota}. A typical DL-based approach involves training an encoder network to produce meaningful representations in a low-dimensional embedding space based on high-dimensional image inputs. To learn this embedding space, the network is trained on `triplets' of data samples: one `anchor' image as the reference, one `positive' image that is either directly paired with the `anchor' image or is semantically relevant to the `anchor' image, and another `negative' image that is irrelevant to the `anchor' image. The network is trained to produce similar embeddings (with a low analytical distance in the embedding space) for the `positive' and `anchor' images and dissimilar embeddings for the `negative' and `anchor' images. When retrieving images with a sketch query, the natural images are ranked by the distance (e.g., Euclidean Distance) between their embedding outputs and the sketch query's embedding outputs. A few of the recently introduced datasets mentioned in Section \ref{sec:datasets} have introduced DL-based architectures that establish state-of-the-art performance over non-neural baselines. 




\subsection{Neural Sketch Generation}
Recent advancements in the DL community have introduced deep neural networks capable of recognizing and generating sketches. Sketch-RNN~\citep{sketchrnn} is one of the first RNN-based models that can generate sequential sketch strokes through supervised \crd{learning} on sketch datasets. We will describe this architecture in detail in Section~\ref{sec:scones_sys} due to its high relevance to our work. Transformer networks~\citep{transformer:vaswani:2017} have shown superior performance in natural language modeling/generation tasks and have recently been applied to model sketch strokes and SVG graphics. These approaches have been shown to improve sketch generation performance over Sketch-RNN. Most recently, CoSE is a hierarchical architecture that aims to improve performance of DL models in generating structured sketches previously difficult for flat models~\citep{cose}. In particular, this approach targets the generation of structured diagrams in the DiDi dataset.

Another highly relevant model architecture is Generative Adversarial Networks (GANs). GANs have also been used to translate realistic images into sketches (or edges) at the pixel level by training on \crd{paired~\citep{photo-sketching}} and unpaired~\citep{cyclegan} sketch and image data. While the pixel-based architectures have been effective when handling natural image data, it does not naturally couple with humans' sketching processes, which are inherently sequential.




\section{Developing a Paired Sketch/User Interface Dataset}
\label{sec:swire_data}
In this section, we explore the first step towards developing a novel deep-learning-driven creativity support application: collecting a large-scale sketch dataset. We specifically target the task of drawing correspondance between two types of visual artifacts commonly used in the early-stage of UI design: low-fidelity design sketches and high-fidelity UI examples. 

Both Sketches and UI design examples are commonly used in the UI design process as reported by a variety of prior studies \citep{newman, bailey} and our informal conversation with designers. Designers search, consult and curate design examples to gain inspiration, explore viable alternatives and form the basis for comparative evaluations \citep{bailey, creativity}. Similarly, designers frequently use sketches to expand novel ideas, visualize abstract concepts, and rapidly compare alternatives \citep{buxton}. As such, understanding correspondences between these two modalities would allow machines to rapidly retrieve popular visual illustrations, common flow patterns and high-fidelity layout implementations \citep{erica} from large corpuses, which can greatly augment various design tasks \citep{bricolage, rewire, remaui}. 

To solve this task using deep-learning-based approaches, we  decided to collect a dataset of actual sketches stylistically and semantically similar to designers' sketches of UIs. This also allows us to leverage large-scale UI datasets recently introduced by mobile-interaction mining applications~\citep{erica}, such that we would only need to collect sketches newly created by designers based on screenshots of original UIs in the Rico dataset~\citep{rico}. The dataset is now publicly available at \url{https://github.com/huang4fstudio/swire}. 

\subsection{Designer Recruitment and Compensation}
We recruited 4 designers through the freelancing platform Upwork. All designers reported having at least occasional UI/UX design experience and substantial sketching experience. In addition, all designers reported receiving formal training in UI design and degrees in design-related fields. They were compensated 20 USD per hour and worked for 60-73 hours. 

\subsection{Dataset Statistics}
We collected 3702 sketches\footnote{This total of 3702 sketches differs from original Swire publication~\citep{swire}. We discovered that 100 trial sketches from a pilot study were accidentally included in the original stated total and we have corrected the numbers in this chapter.} of 2201 UI examples from 167 popular apps in the Rico dataset. Each sketch was created with pen and paper in 4.1 minutes on average. Many UI examples were sketched by multiple designers. 66.5\% of the examples were sketched by 2 designers, 32.7\% of the examples were sketched by 1 designer and the remaining examples (< 1\%) were sketched by 3 designers in our dataset. Our 4 designers sketched 455/1017/1222/1008 UIs respectively based on their availability. We allocated batches of examples to different combinations of designers to ensure the generality of the dataset.

We did not have the resources to generate sketches for every UI in the Rico dataset, so we curated a diverse subset of well-designed UI examples that cover 23 app categories in the Google Play Store and were of average to high design quality. We omitted poorly designed UIs from the dataset because of the relatively small size of the dataset for neural network training. Noise introduced into training by poor designs would have the potential to negatively impact the training time and quality of our model.

\subsection{Data Collection and Postprocessing Procedure}
\begin{figure}
\centering
  \includegraphics[width=0.9\columnwidth]{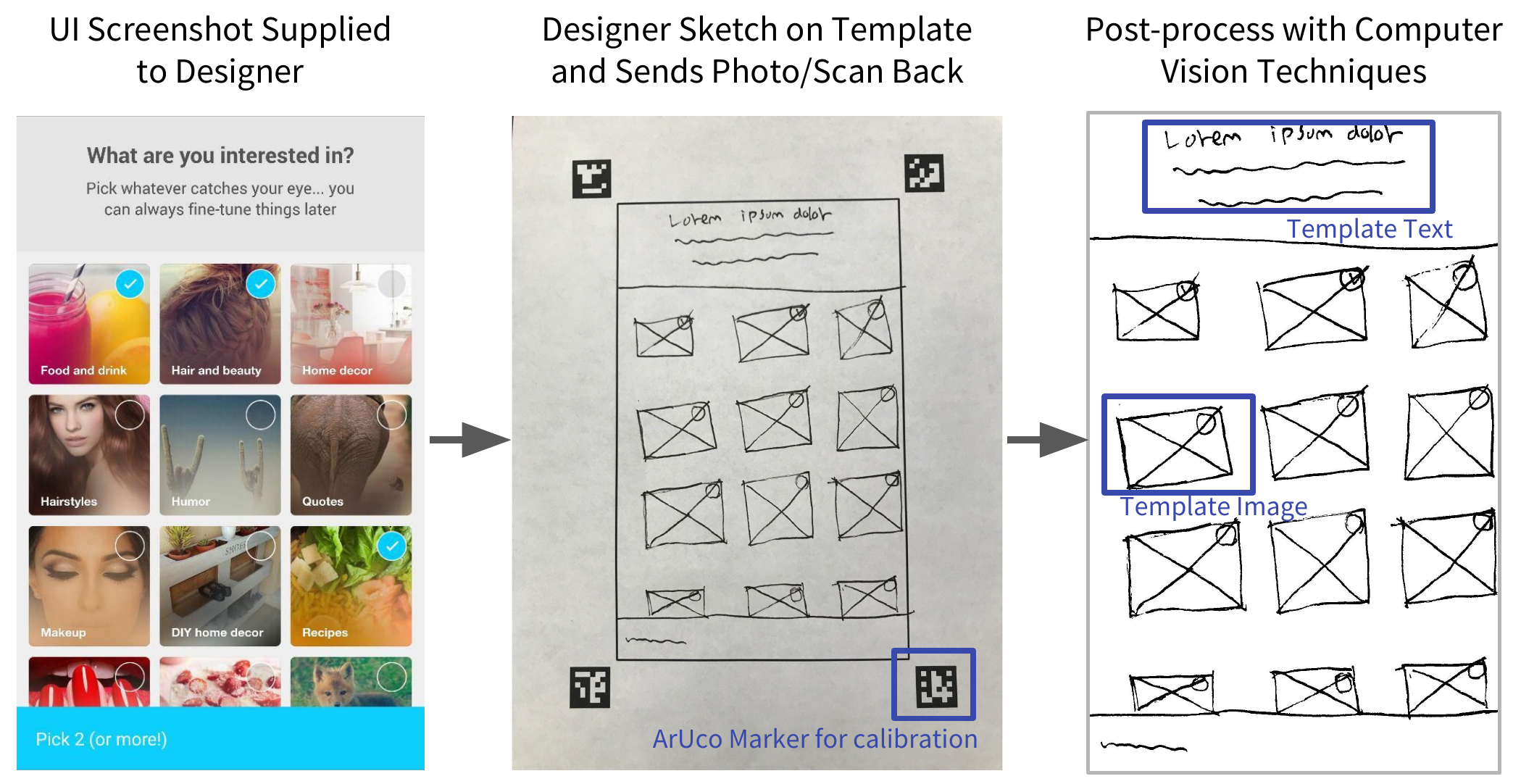}
  \caption{Data Collection Procedure.
  We first send a UI screenshot (left) and paper templates with ArUco markers to designers. Designers then sketch on the template and sends back a photo or a scan of the completed sketch (middle). We then post-process the photo using Computer Vision techniques to obtain the final clean sketch dataset (right). }~\label{fig:aruco}
\end{figure}

We supplied the screenshots of our curated UI examples to the recruited designers and asked them to create sketches corresponding to the screenshots with pen and paper. They were prompted to reconstruct a low-fidelity sketch from each screenshot as if they were the designers of the interfaces. We instructed them to replace all actual image content with a sketched placeholder (a square with a cross or a mountain) and replace dynamic text with template text in each screenshot as shown in Figure \ref{fig:aruco}. We added these instructions to obtain sketches with a more unified representation focused on the design layout of various UIs. These instructions also make it easier for the neural network to learn the concepts of images and text within the constraints of our small dataset. 

In order to efficiently collect and calibrate sketches created by multiple designers in various formats of photos and scans, we supplied them with paper templates with frames for them to sketch on as shown in Figure \ref{fig:aruco}. These frames are annotated with four ArUco codes \citep{aruco} at the corners to allow perspective correction. All photos and scans of the sketches are corrected with affine transformation and thresholded to obtain binary sketches as final examples in the dataset.

\section{Developing Swire: a Sketch-based User Interface Retrieval System}
\label{sec:swire_sys}
After collecting a dataset of corresponding UIs and screenshots, we proceed to develop Swire, a sketch-based user interface retrieval system that learns correspondances between UIs and sketches in the dataset. In developing Swire, we adapted a commonly used DL architecture, training paradigm, and loss function that are able to retrieve relevant visual content given input sketches. Similar to prior work in sketch-based image retrieval, we used a convolutional neural network architecture and adopt the cross-modal embedding training scheme using a triplet loss. This allows us to take advantage of a convolutional neural network's strong capability to understand high-dimensional visual features while creating a unified embedding space for both sketches and UIs with learned concepts based on their correspondences. This means Swire can be used to search a dataset of UIs using either sketches or UI screenshots as the querying modality.

The development of Swire consists of a training phase and a querying phase. During the training phase, we train Swire's deep neural network to generate similar low-dimensional outputs (64-dimensional) for matching pairs of screenshots and sketches, and dissimilar outputs for non-matching pairs of screenshots and sketches. This training scheme is shown to be useful for sketch-based image retrieval \citep{sketchy}. In the querying phase, we use Swire's trained neural network to encode a user's sketch query and retrieve UIs with the closest output to the sketch query's output.

\subsection{Network Architecture}
\begin{figure*}
\centering
  \includegraphics[width=0.9\textwidth]{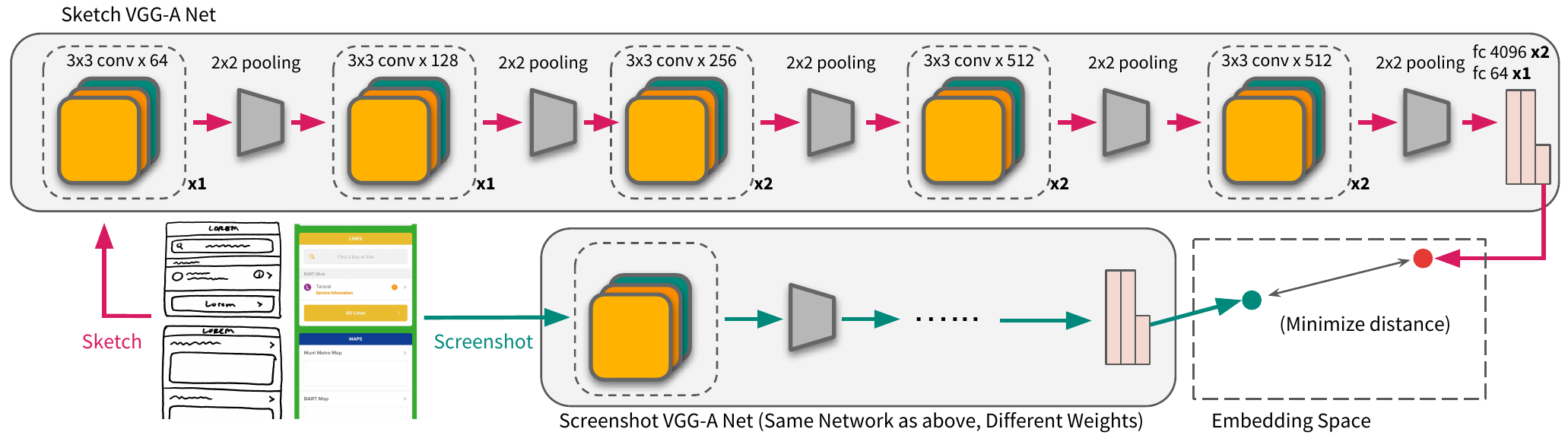}
  \caption{Network Architecture of Swire's Neural Network. Swire's neural network consists of two identical sub-networks similar to the VGG-A deep convolutional neural network. These networks have different weights and attempts to encode matching pairs of screenshots and sketches with similar values.}~\label{fig:network}
\end{figure*}

We used two convolutional sub-networks to handle the two inputs of sketch-screenshot pairs, these two sub-networks are similar to VGG-A \citep{vgg}, a shallow variant of the state-of-the-art network that won the ILSVRC2014 image recognition challenge \citep{ilsvrc2014}. Our network consists of 11 layers, with five convolutional blocks and three fully-connected layers. Each convolutional block contains two (one for the first two blocks) convolutional layers with 3x3 kernels and one max-pooling layer. The convolutional layers in the five blocks have 64, 128, 256, 512, and 512 filters respectively. The first two fully-connected layers have 4096 hidden units. The last layer has 64 hidden units and outputs the 64-dimensional embedding used for querying. The activation functions of all layers except the last layer are ReLU. The network architecture is described in detail in Figure \ref{fig:network}.

The final 64-dimensional output embeddings of the sub-networks are trained to produce adequate embeddings represented as codes in their respective final layers. The model is trained with a pairwise sampling scheme described in the following subsection.

\subsection{Triplet Loss}
\label{sec:loss}
The model is trained with a Triplet Loss function \citep{triplet, smts} that involves the neural-network outputs of three inputs: an `anchor' sketch $s$, a `positive' matching screenshot $i$ and a `negative' mismatched screenshot $i'$. This forms two pairs of input during training. The positive pair $p(s, i)^+$ consists of a sketch-screenshot pair that correspond to each other. The negative pair $p(s, i')^-$ consists of a sketch-screenshot pair that does not correspond. The negative pair is obtained with the same sketch from the positive pair and a random screenshot sampled from the mini-batch. 

During training, each pair $p(s,i)$ is passed through two sub-networks such that the sketch sample $s$ is passed through the sketch sub-network and outputs an embedding $f_s(s)$, and we similarly obtain the neural-network output of the screenshot $f_i(i)$ from the screenshot sub-network. We compute the Euclidean distance $D$ between the neural network outputs. For the positive pair, 
$$D(p(s, i)^+) = ||f_s(s) - f_i(i)||_2$$
Similarly, for the distance of the negative pair, $$D(p(s, i')^-) = ||f_s(s) - f_i(i')||_2$$
With these distances, we formulate a triplet loss function,
$$L = D(p(s, i)^+) + \max{(0, m - D(p(s, i')^-))}$$
$$ m = \text{margin between positive and negative pairs}$$

We maintain a margin $m$ between the positive and negative pairs to prevent the network from learning trivial solutions (zero embeddings for all examples).

\subsection{Data and Training Procedure}
We train our network using the dataset described in Chapter~\ref{sec:swire_data}. Since the sketches are created by four separate designers, we split the data and used data collected from three designers for training and from one designer for testing. This is to ensure that the model generalizes across sketches produced by different designers. In addition, we do not repeat interfaces from the same apps between the training and test sets. This creates 1722 matching sketch-screenshot pairs for training and 276 pairs for testing.

During training, the sketches and screenshots are resized to $224 \times 224$ pixels, and the pixel values are normalized between $(-1, 1)$ centered at $0$. The network is trained using a Stochastic Gradient Descent Optimizer with a mini-batch size of 32. The learning rate is $1 \times 10^{-2}$. The margin is $0.2$ in all models. All hyper-parameters listed above were determined by empirical experiments on the training set.

\subsection{Querying}
When a user makes a query with a drawn sketch, the model computes an output by passing the sketch through the sketch sub-network. This output is then compared with all neural-network outputs of the screenshots of UI examples in the dataset using a nearest neighbor search. The UI results are ranked by the distance between their outputs and the user's sketch's output. 

\subsection{Results}
\label{sec:results}
\subsubsection{Baselines}
We implement competitive non-neural baselines to evaluate the performance of our method. As described in Section \ref{sec:sbir}, typical methods of sketch-based image retrieval involve two steps: (1) extract an edge-map from the original image to be queried, and (2) match the edge-map using a specific similarity metric. Using this framework, we first extracted the edges of the screenshots using the Canny Edge detector \citep{canny}. We then extracted features from the edges using Bag-of-words (BoW) Histogram of Oriented Gradients (HOG) filters. BoW-HOG filters is an advanced method of computing similarity between images. They capture edge features in an image by computing the magnitude of gradients across the entire image with respect to multiple orientations. This method summarizes image features with fixed-length vectors that describe the occurrences and characteristics of edges in images. This method is highly effective for sketch-based image retrieval as it focuses on the characteristics of edges while being insensitive to local translations and rotations. 

After obtaining these fixed-length vectors for screenshots and sketch queries, we compare them using Euclidean distance as a simple metric to query for closest matching images (design screenshots in our case) to the sketch queries. 

\subsubsection{Quantitative Results}
We use a test set that consists of 276 UI examples to compare Top-1 and Top-10 performances of BoW-HOG filters and Swire. The results are summarized in Table \ref{table:quant}. We observe that Swire significantly outperform BoW-HOG filters for Top-10 performance at 60.9\%. For Top-1 accuracy, Swire achieves an accuracy of 15.9\% which only slightly outperformed the strong baseline of BoW-HOG filters at 15.6\%. This shows Swire to be particularly effective for retrieving complex examples from the dataset compared to the BoW-HOG filters. We believe deep-learning-based Swire is advantageous compared to BoW-HOG filters that rely on matching edge-maps because UI sketches have semantic complexities that are not captured by edge-maps of screenshots. 

\begin{table}
  \centering
  \begin{tabular}{l r r}
    {\small\textbf{Technique}}
    & {\small \textbf{Top-1}}
      & {\small \textbf{Top-10}} \\
    \midrule
    (Chance) & 0.362\% & 3.62\% \\ 
    BoW-HOG filters & 15.6\% & 38.8\% \\
    \textbf{Swire} & \textbf{15.9\%} & \textbf{60.9\%} \\
  \end{tabular}
  \caption{Top-k Accuracy of Various Models on the Test Set. Swire significantly outperforms BoW-HOG filters.}~\label{table:quant}
\end{table}

\subsubsection{Qualitative Results}
\begin{figure}
\centering
  \includegraphics[width=1.0\columnwidth]{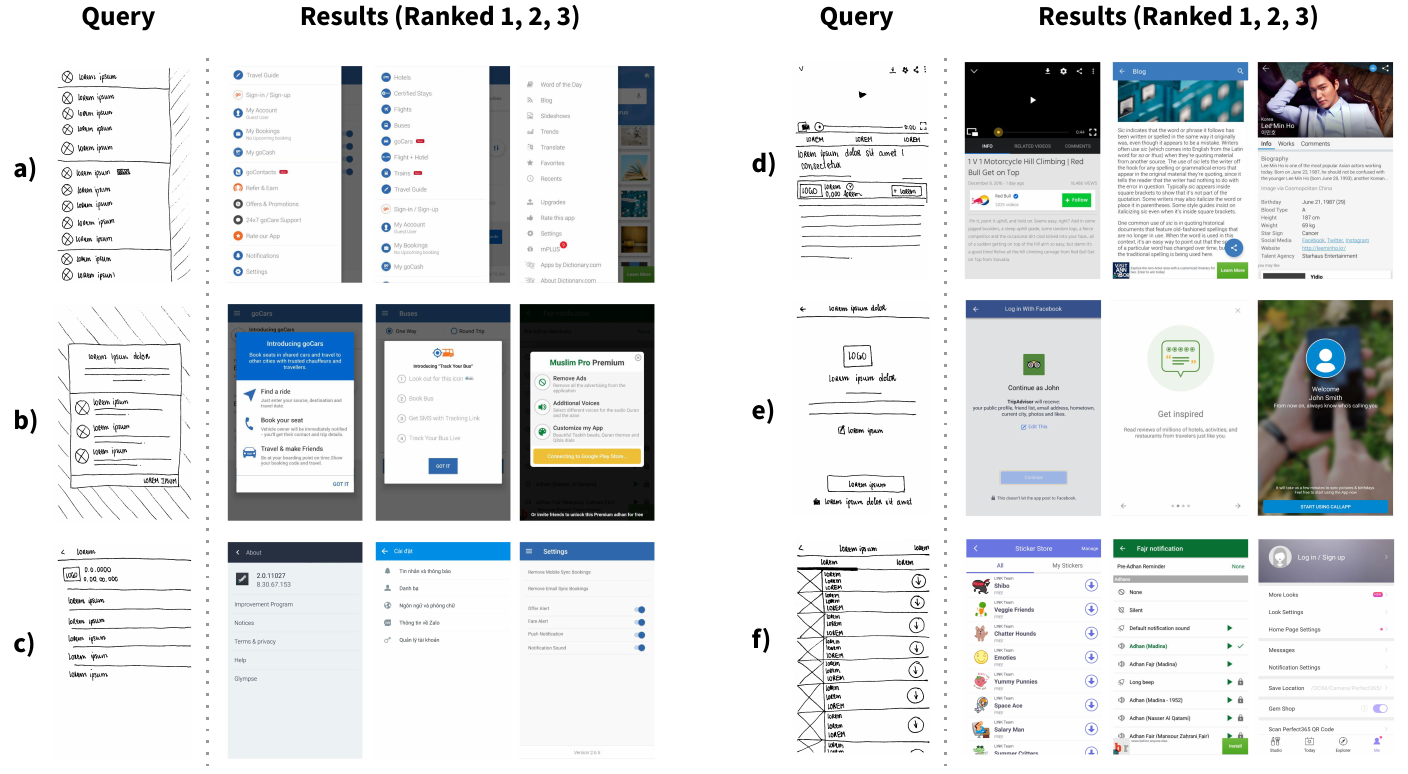}
  \caption{Query Results for Complete Sketches. Swire is able to retrieve common types of UIs such as sliding menus (a), settings (c), and login (e) layouts.}~\label{fig:query_main}
\end{figure}

We visualize query results from the test set to qualitatively understand the performance of Swire in Figure \ref{fig:query_main}. Swire is able to retrieve relevant menu-based interfaces (Example a) despite the difference in visual appearance of the menu items. Swire is also able to retrieve pop-up windows (Example b) implemented in various ways despite the drastic difference in the dimensions of the pop-up windows. We observe similar efficacy in retrieving settings (Example c), list-based layouts (Example f), and login layouts (Example e). Nevertheless, we observe that Swire sometimes ignores smaller details of the interfaces described by sketched elements. This limitation will be further discussed in Section \ref{sec:future_data}.

\subsubsection{Expert Evaluation}
To better evaluate Swire's performance from professional users' perspectives, we recruited 5 designers on Upwork with substantial experience in mobile UI/UX design to evaluate selected results from the test set. There was no overlap between these designers and those recruited for creating the dataset. We provided them with 9 sets of query sketches and the corresponding Top-10 retrieved results for each query from the test set. The 9 sets consist of 3 `best' results (the corresponding screenshot of the sketch query is retrieved as the Top-1 result), 3 `mediocre' results (the corresponding screenshot of the sketch query is retrieved within the Top-10 results, but not Top-1), and 3 `poor' results (the corresponding screenshot of the sketch query is not retrieved within the Top-10 results). We asked the designers to provide comments on each set of results regarding the relevance between the sketches and the screenshots, and to comment on the potential integration of this tool into their design workflows.

Most designers agreed that all retrieved results in the `best' result sets are relevant to the query, and they would be satisfied with the results. They were especially satisfied with a result set of sliding menus (also shown in Figure \ref{fig:query_main}a). They were able to identify the results as `variations on the theme of navigation drawers' (D3) or `slide out modal pattern.' (D2) Moreover, the designers also expressed satisfaction towards some sets of `mediocre' results. Most were satisfied with a set of results that `show variations of the top tabbed navigation' (D5) which is a common design pattern.

On the other hand, some designers considered the `poor' results unsatisfactory. For example, designers were less satisfied with the model's performance on a sign-up sketch, commenting that the model only gathered screens with similar element layouts while ignoring the true nature of the input fields and buttons in the query (D3). However, D4 considered `rows of design elements' common in the results relevant to the sketch, and D1 considered two similar sign-up screens retrieved by the model as strong results even they did not match up perfectly with the sketch.

In general, we observed that designers were more satisfied with the results when the model was able to retrieve results that are semantically similar at a high-level instead of those with matching low-level element layouts. Notably, D1 commented that we `probably already considered the common UI sketch patterns and train[ed] [our] system to match it up with image results,' which reflects the effectiveness of Swire in detecting common UI patterns in some instances provided that it was not specifically trained to recognize these patterns. All designers also considered Swire to be potentially useful in their workflows for researching, ideating and implementing novel designs.

\subsection{Applications}
In Section \ref{sec:results}, we evaluated and validated Swire's effectiveness for generally finding design examples through sketch-based queries. Since both sketches and UI design examples are commonly used in early stages of the UI design process as reported by a variety of prior studies \citep{newman, bailey}, we explore the potential usage of Swire through several design applications in this section. Prototypes of these applications implemented with Jupyter Notebook are available at \url{https://github.com/huang4fstudio/swire}.

\subsubsection{Evaluation with Alternative Designs}
Designers often explore alternative design examples to support the implementation and comparative evaluation \citep{bailey} of their own designs. HCI research literature also recommends the use of parallel prototyping techniques to obtain better final products through extensive comparison \citep{dow}. Swire is able to support design comparisons because it enables querying for similar UIs with high-fidelity UI prototypes.

Swire is effective in retrieving similar UIs because the visual content of UI screenshots are reinforced with the semantic structure of sketches in the embedding space during training. Swire can thus be used as a semantically-aware similarity metric between interfaces. 

\begin{figure}
\centering
  \includegraphics[width=0.5\columnwidth]{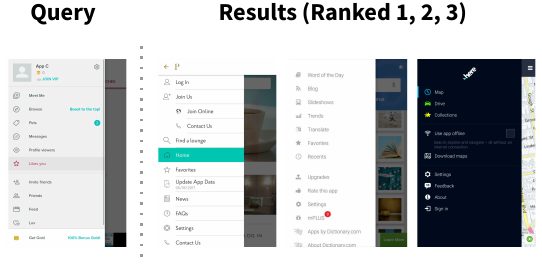}
  \caption{Alternative Design Query Results. Swire is able to retrieve similar UIs in the dataset from queries of complete, high-fidelity UI screenshots.}~\label{fig:ui}
\end{figure}

Figure \ref{fig:ui} shows an example of Swire retrieving similar menus using a high-fidelity screenshot. We similarly observe that it is able to retrieve design patterns such as login screens, list-based UIs, and grid-based UIs when querying with screenshots (please see the original Swire paper~\citep{swire} for a complete figure illustrating these patterns). This enables effective comparison between similar designs with slight variations. 

\subsubsection{Auto-completing Partial Designs}
Sketches are often used for rapid exploration of potential design solutions \citep{buxton}. Designers use partial sketches to express core ideas, while leaving out parts of the interface in sketches for considering viable design alternatives. We trained an alternative model \emph{Swire-segments} on partial sketches of UIs, which allows us to `auto-complete' the remaining UI by retrieving a variety of examples that are only required to match parts of the UI sketched by the user. This model allows designers to quickly gain design inspirations that are relevant to the key UI elements desired by them.

In the training and querying phases of \emph{Swire-segments}, UI examples are split into small parts. Designers can thus specify one-or-more parts of the UIs to be matched by the model with examples in the dataset. We compute an embedding for each part of the interface and match only the embeddings of the parts specified by the users for retrieval. Figure~\ref{fig:autocomplete} shows that Swire-segments is able to retrieve multiple designs that all contain the Floating Action Button (FAB, a popular Android design paradigm) but with diverse layouts. Swire-segments is also able to retrieve interfaces with only tab-based top bars in common. In these examples, Swire-segments is able to remain agnostic to the unspecified part of the sketch queries.

\subsubsection{User Flow Examples}
Beyond querying for single UIs, designers also use sketches to illustrate user experience at multiple scales \citep{newman}, such as conveying transitions and animations between multiple interfaces. Since the Rico dataset also includes user interaction data, we use this data to enable flow querying with Swire. Designers can use this application to interact with interaction design examples that can accelerate the design of effective user flows. 

\begin{figure*}[htp] 
    \centering
    \subfloat[Autocomplete Query Results]{%
        \includegraphics[width=0.48\columnwidth, valign=b]{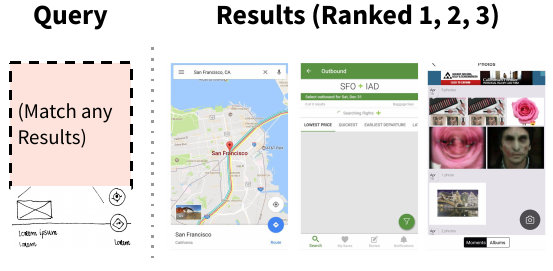}%
        \label{fig:autocomplete}%
        }%
    \hfill%
    \subfloat[Flow Query Results]{%
        \includegraphics[width=0.48\columnwidth, valign=b]{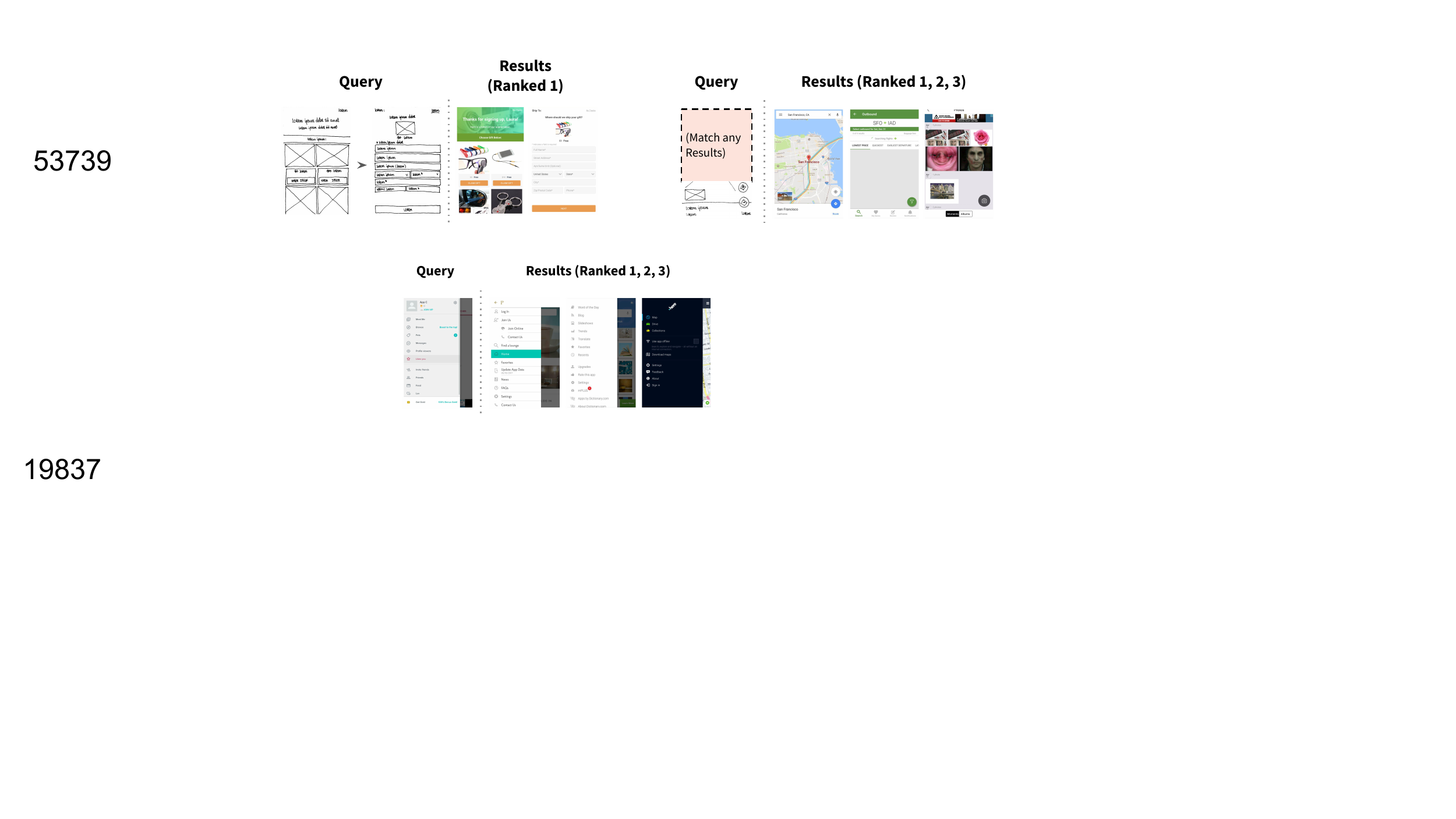}%
        \label{fig:flow}%
        }%
    \caption{Query Results for (a) incomplete sketches and (b) flow queries. Swire is able to retrieve interfaces only based on parts specifed by users’ sketches while remaining agnostic to other parts of the UIs. Swire is also able to retrieve user flows by querying multiple UIs in sequences concurrently.}
\end{figure*}

To query flow examples in the dataset, since Swire creates a single embedding for each UI, we can match an arbitrary number of interfaces in arbitrary order by concatenating the embedding values during the ranking process of querying. We qualitatively observe that Swire is able to retrieve registration (Figure~\ref{fig:flow}) and `closing menu' flows that are commonly implemented by designers. Since Rico also contain transition details between consequent UIs, these examples can demonstrate popular animation patterns \citep{erica} that provide inspiration to interaction and animation designers.

\section{Developing Scones: a Conversational Sketching System}
\label{sec:scones_sys}
Beyond comprehending and understanding the correspondances between UIs and sketches, in this section we explore the possibility of sketch generation through another more abstract input modality: natural language. 
In particular we develop a system that supports {\em iterative generation} of sketches given users' text instructions to support refinement and critique. The goal is to mimic a user trying to verbally convey a visual idea to an expert sketcher. 

While sketching is a powerful communication medium, creating sketches that effectively communicate ideas visually requires significant training. The use of sketches in an iterative design and/or artistic process, where the sketch itself is annotated or refined, requires additional, specialized expertise. Inspired by recent development of deep-learning-driven QA systems and generative sketching models, we develop Scones, a deep-learning-based sketching system that can progressively construct a sketched scene based on multiple natural language instructions across multiple turns, an interaction analogous to an iterative sketch/critique process. This system must unify knowledge of the \textit{low-level} mechanics for generating sketch strokes and natural language modification instructions with a \textit{high-level} understanding of composition and object relationships in scenes.

We formulate the novel task of iteratively generating and refining sketches with text instructions and present a web-deployable implementation of \systemnamenospace. \systemname contains a scene composition proposer that takes a novel approach in creating and editing scenes of objects using natural language. It adapts a recent neural network architecture and improves state-of-the-art performance on the scene modification task. We also introduce in \systemname a novel method for specifying high-level scene semantics within individual object sketches by conditioning sketch generation with mask outlines of target sketches. Using Scones, we hope to enable users of all levels of sketch expertise to freely express their intent using abstract, text-based instructions, together with concrete visual media.

\subsection{System Architecture}
\begin{figure}[h]
  \centering
  \includegraphics[width=\linewidth]{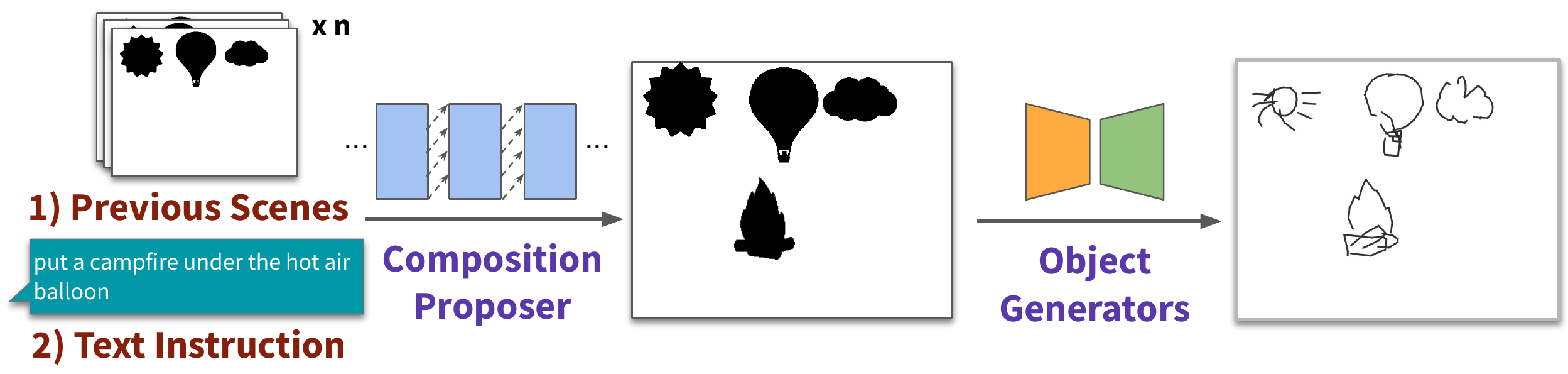}
  \caption{Overall Architecture of Scones. Scones takes a two-stage approach towards generating and modifying sketched scenes based on users' instructions.}
  \label{fig:sys}
\end{figure}

The creation of complex sketches often begins with semantic planning of scene objects. Sketchers often construct high-level scene layouts before filling in low-level details.
Modeling ML systems after this high-to-low-level workflow has been shown to be beneficial for transfer learning from other visual domains and for \crd{supporting} interactive interfaces for human users~\citep{sketchforme}.
Inspired by this high-to-low-level process, \systemname adopts a hierarchical workflow that first proposes a \crd{scene-level} composition \crd{layout} of objects using its \emph{Composition Proposer}, then generates individual object sketches, conditioned on the scene-level information, using its \emph{Object Generators} (Figure~\ref{fig:sys}).

\subsubsection{Composition Proposer}
The Composition Proposer in \systemname uses text instructions to place and configure objects in the scene. It also considers recent past iterations of text instructions and scene \emph{context} at each conversation turn. As text instructions and sketch components occur sequentially in time, each with a variable length of tokens and objects, respectively, we formulate composition proposal as a sequence modeling task. We use a self-attention-only decoder component of the Transformer~\citep{transformer:vaswani:2017}, a recent deep-learning model architecture with high performance for this task.

To produce the output scene $S_i$ at turn $i$, the Composition Proposer takes inputs of $n = 10$ previous scenes $S_{(i - n), \dots ,(i - 1)}$ and text instructions $C_{(i-n), \dots , (i - 1)}$ as recent \emph{context} of the conversation. Each output scene $S_i$ contains $l_i$ objects $o_{(i, 1), \dots , (i, l_i)} \in S_i $ and special tokens $o_s$ marking the beginning and $o_e$ marking the end of the scene. Each text instruction $C_i$ contains $m_i$ text tokens $t_{(i, 1), \dots , (i, m_i)} \in C_i$ that consist of words and punctuation marks. 

We represent each object $o$ as a 102-dimensional vector $o$: 
$$o = [\mathbbm{1}_{s}, \mathbbm{1}_{e}, e^{(o)}, e^{(u)}, e^{(s)}, e^{(f)}, x, y]$$
The first two dimensions $\mathbbm{1}_{s}, \mathbbm{1}_{e}$ are Boolean attributes reserved for the start and end of the scene object sequences. $e^{(o)}$ is a 58-dimensional one-hot vector\sig{\footnote{\sig{an encoding of class information that is an array of bits where only the corresponding position for the class to be encoded is 1, and all other bits are 0s.}}} representing one of 58 classes of the scene objects $e^{(u)}$ is a 35-dimensional one-hot vector representing one of 35 sub-types (minor variants) of the scene object. $e^{(s)}$ is a three-dimensional one-hot vector representing one of three sizes of the scene objects. $e^{(f)}$ is a two-dimensional one-hot vector representing the horizontal orientation of whether the object is flipped in the x-direction. The last two dimensions $x, y \in [0, 1]$ represents the x and y position of \sig{the center of the object}. This representation is very similar to that of the CoDraw dataset the model was trained on, which is described in detail in Section \ref{sec:codraw}. For each text token $t$, we use a 300-dimensional GLoVe vector trained on 42B tokens from the Common Crawl dataset~\citep{glove} to semantically represent these words \crd{in} the instructions.

To train the Transformer network with the \crd{heterogeneous} inputs of $o$ and $t$ across the two modalities, we create a unified representation of cardinality $|o| + |t| = 402$  and adapt $o$ and $t$ to this representation by simply padding additional dimensions in the representations with zeros as shown in Equation \ref{eqn:padding1}.

\begin{equation}
\label{eqn:padding1}
    o'_{i, j} = [o_{i, j}, \Vec{0}_{(300)}] \quad
    t'_{i, j} = [\Vec{0}_{(102)}, t_{i, j}]
\end{equation}
 
\begin{figure}[h]
  \centering
  \includegraphics[width=\linewidth]{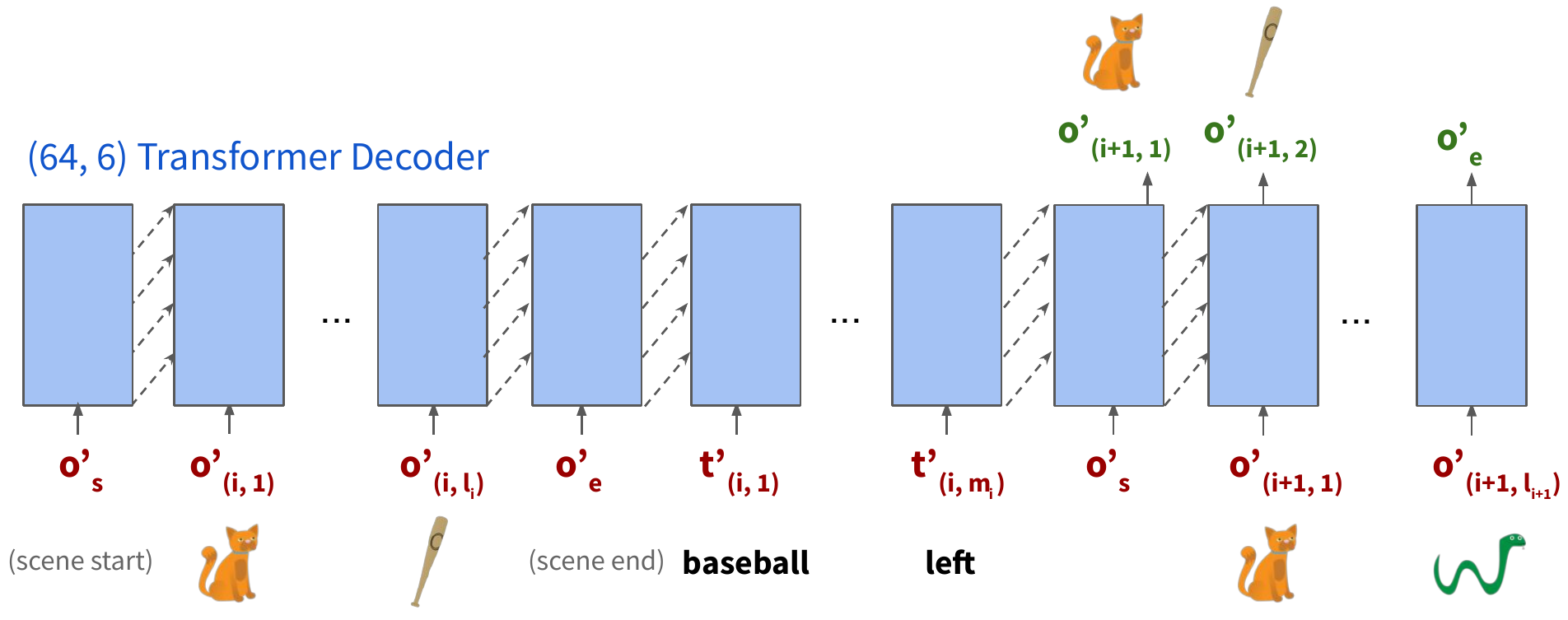}
  \caption{The Scene Layout Generation Process using the Transformer Model of the \textit{Composition Proposer.}}
  \label{fig:transformer}
\end{figure}

We interleave text instructions and scene objects chronologically to form a long sequence $[C_{(i-n)}, S_{(i-n)}$ $, ... , C_{(i-1)}, S_{(i-1)}, C_i]$ as input to the model for generating an output scene representation $S_{i}$. We expand the sequential elements within $C$ and $S$, and add separators to them to obtain the full input sequence to \crd{a single} Transformer Decoder. To adapt the Transformer model to our multi-modal inputs $t'$ and $o'$ and produce new scene objects $o$, we employ a 402-dimensional input embedding layer and 102-dimensional output embedding layer \crd{in the Transformer model}. The outputs from the network are then passed to sigmoid and softmax activations for object position and other properties respectively. We show this generation process in Equation \ref{eqn:transformer} and in Figure \ref{fig:transformer}.


\begin{multline}
\label{eqn:transformer}
    S_i = [o_{(i, 1), ... , (i, l)}] = 
    \textbf{Transformer}([o'_s, o'_{(i - n, 1)}, ... o'_{(i - n, l_{(i-n)})}, \\ o'_e, t'_{(i - n, 1)}, ... , t'_{(i - n, m_{(i - n)})}, ... , t'_{(i, 1)}, ... t'_{(i, l_i)}, o'_s])
\end{multline}

\label{sec:object-sketchers}
\subsubsection{Object Generators}
Since the outputs of the Composition Proposer are \crd{scene layouts consisting of} high-level object \crd{specifications}, we generate the final raw sketch strokes for each of these objects based on their \crd{specifications} with \emph{Object Generators}. We adapt Sketch-RNN to generate sketches of individual object classes to present to users for evaluation and revision in the next conversation turn. Each sketched object $Q$ consists of $h$ strokes $q_{1 \dots h}$. The strokes are encoded using the \emph{Stroke-5} format~\citep{sketchrnn}. Each stroke $q = [\Delta x, \Delta y, p_d, p_u, p_e]$ represents states of a pen performing the sketching process. The first two properties $\Delta x$ and $\Delta y$ are offsets from the previous point that the pen moved from. The last three elements $[p_d, p_u, p_e]$ are a one-hot vector representing the state of the pen after the current point (pen down, pen up, end of sketch, respectively).
All sketches begin with the initial stroke $q_1 = [0, 0, 1, 0, 0]$.


Since Sketch-RNN does not constrain aspect ratio\crd{s}, direction\crd{s} and pose\crd{s} of its output sketches, we introduce two additional conditions for the sketch generation process: masks $m$ and aspect ratios $r$. These conditions ensure our Object Generators generate sketches with appearances that follow the object specifications generated by the Composition Proposer. For each object \crd{sketch}, we compute the aspect ratio $r = \dfrac{\Delta y}{\Delta x}$ by taking the distance between the leftmost and rightmost stroke as $\Delta x$ and the distance between topmost and bottommost stroke as $\Delta y$.
To compute the object mask $m$, we first render the strokes into a pixel bitmap, then mark all pixels as 1 if they are in between the leftmost pixel $py_{xmin}$ and rightmost pixel $py_{xmax}$ that \crd{are passed through by} any strokes for each row $y$, or if they are in between the bottommost pixel $px_{ymin}$ and topmost pixel $px_{ymax}$ that \crd{are passed through by} any strokes for each column $x$ (Equation~\ref{eqn:mask}). As this mask-building algorithm only involves pixel computations, we can use the same method to build masks for clip art objects (used to train the Composition Proposer) to generate sketches with poses matching the Composition Proposer's object representations.

\begin{equation}
\label{eqn:mask}
    m_{(x, y)} =
\left\{
	\begin{array}{ll}
		1  & \mbox{if } py_{xmax} \geq x \geq py_{xmin}, \mbox{or}; \\
		1 &\mbox{if } px_{ymax} \geq y \geq px_{ymin} \\
		0 & \mbox{otherwise}
	\end{array}
\right.
\end{equation}

\begin{figure}[h]
  \centering
  \includegraphics[width=0.8\linewidth]{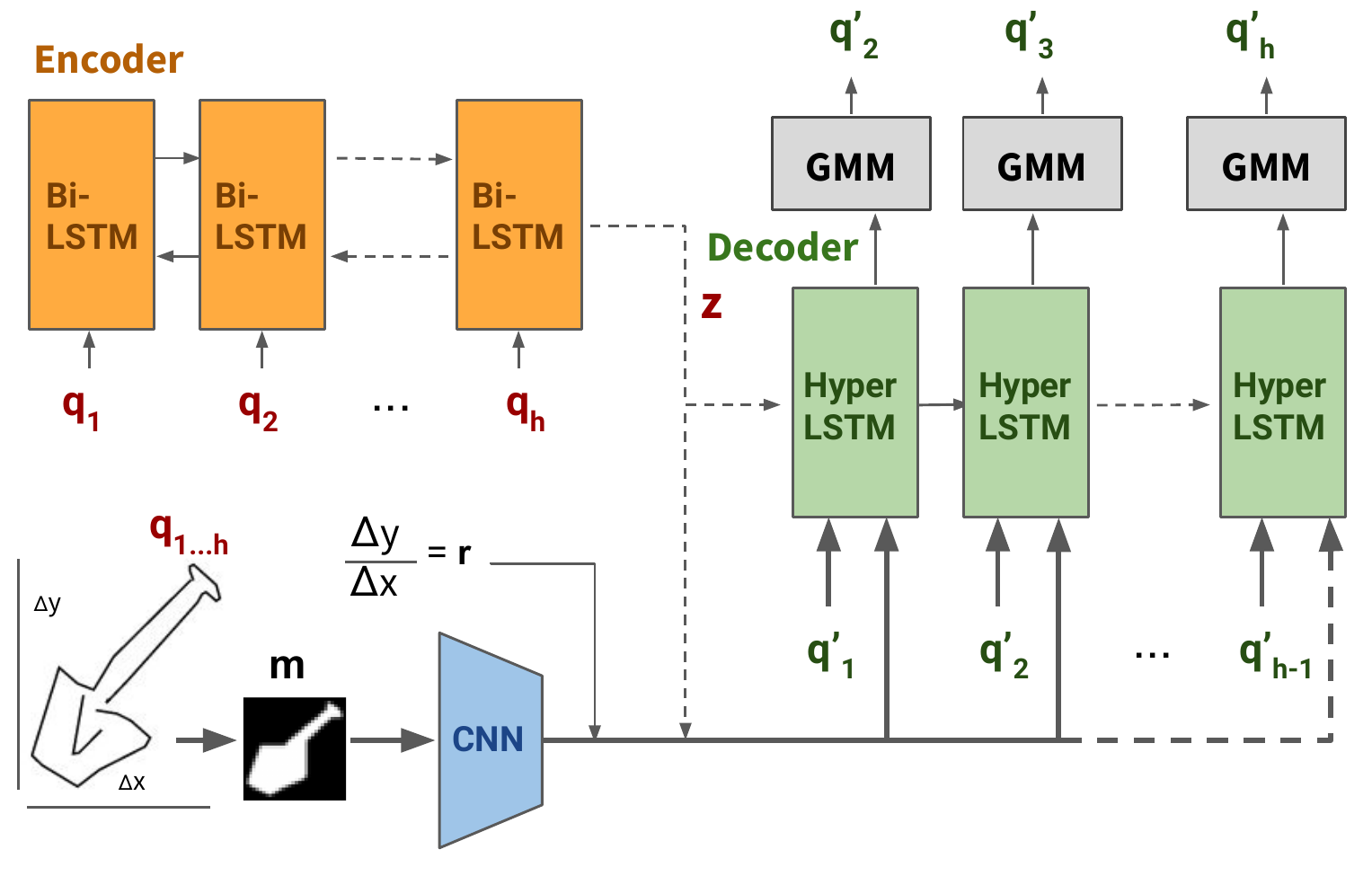}
  \caption{Sketch-RNN Model Architecture of the Object Generators.}
  \label{fig:sketch-rnn}
\end{figure}

We adapt the Variational-Autoencoder(VAE)-based conditional variant of Sketch-RNN to enable generating and editing of sketch objects. Our adapted conditional Sketch-RNN encodes input sketches with a Bi-directional LSTM to a latent vector $z$. The Hyper-LSTM decoder then recreates sketch strokes $q'_{1 \dots h}$ from $z$, and $m, r$ described above during training, as defined in Equation \ref{eqn:sketch-rnn} and shown in Figure \ref{fig:sketch-rnn}. Since the latent space is also trained to match a multivariate Gaussian distribution, the Object Generators can support sketch generation when the objects are first added to the scene by randomly sampling $z \sim N(0, 1)^{128}$.

\begin{multline}
\label{eqn:sketch-rnn} 
q'_{1 \dots h} = \textbf{Sketch-RNN Decoder}([m, r, z]), z \sim N(0, 1)^{128} \\
z = \textbf{Sketch-RNN Encoder}(q_{1 \dots h})
\end{multline}

As $m$ is a two-dimensional mask, we encode $m$ using a small convolutional neural network into a flattened embedding to be concatenated with $z$, $r$ and $q_i$ as inputs to the decoder. The decoder then outputs parameters for a Gaussian Mixture Model (GMM) which will be sampled to obtain $\Delta x$ and $\Delta y$. It also outputs probabilities for a categorical distribution that will be sampled to obtain $p_d, p_u$ and $p_e$. This generation process and the architecture of the model are illustrated in Figure \ref{fig:sketch-rnn}, and are described in the Sketch-RNN paper~\citep{sketchrnn}.

\subsection{Datasets and Model Training}
As \systemname uses two components to generate scenes of sketched objects, it is trained on two datasets that correspond to the tasks these components perform.

\subsubsection{CoDraw Dataset}
\label{sec:codraw}
We used the CoDraw dataset~\citep{codraw} to train the Composition Proposer to generate high-level scene \crd{layout} proposals from text instructions. The task used to collect this data involves two human users taking on the roles of \emph{Drawer} and \emph{Teller} in each session. First, the Teller is presented with an abstract scene containing multiple \crd{clip art} objects in certain configurations, and the Drawer is given a blank canvas. The Teller provides instructions using only text in a chat interface to instruct the Drawer \crd{on} how to modify clip art \crd{objects} in the scene. The Teller has no access to the Drawer's canvas in most conversation turns, except in one of the turns when they can decide to `peek' at the Drawer's canvas. 
The dataset consists of 9993 sessions of conversation records, scene modifications, and ground-truth scenes.  

Using this dataset, we trained the Composition Proposer to respond to user\crd{s'} instructions given past instructions and scenes. We used the same training/validation/test split \crd{as} the original dataset.
Our model is trained to optimize the loss function $L_{cm}$ that corresponds to various attributes of the scene objects in the training set: 

\begin{equation}
L_{cm} = L_{c} + \lambda_{\textbf{sub}} L_{\textbf{sub}} + \lambda_{\textbf{flip}} L_{\textbf{flip}} + \lambda_{\textbf{size}} L_{\textbf{size}} + \lambda_{xy} L_{xy}    
\end{equation}

$L_{c}$ is the cross-entropy loss between the one-hot vector of the true class label and the predicted output probabilities by the model. Similarly $L_{\textbf{flip}}$ and $L_{\textbf{size}}$ are cross-entropy losses for the horizontal orientation and size of the object. $L_{xy}$ is the Euclidean Distance between predicted position and true position of the scene object. We trained the model using an Adam Optimizer with the learning rate of $lr = 1 \times 10^{-4}$ for 200 epochs. We set $\lambda_{\textbf{sub}} = 5.0 \times 10^{-2}$, $\lambda_{\textbf{flip}} = 5.0 \times 10^{-2}$, $\lambda_{\textbf{size}} = 5.0 \times 10^{-2}$, $\lambda_{xy} = 1.0$. These hyper-parameters were tuned based on empirical experiments on the validation split of the dataset. 

\subsubsection{Quick, Draw! Dataset}
The Quick, Draw! dataset consists of sketch strokes of 345 concept categories created by human users in a game in 20 seconds~\citep{quickdraw}. We trained our \crd{34} Object Generators on 34 categories of Quick, Draw! data to create sketches of individual objects.

Each sketch stroke in Quick, Draw! was first converted to the Stroke-5 format. $\Delta x$s and $\Delta y$s of the sketch strokes were normalized with their standard deviations for all sketches in their respective categories. Each category consists of 75000/2500/2500 sketches in the training/validation/test set. 

The loss function of the conditional Sketch-RNN $L_{s}$ consists of the reconstruction loss $L_R$ and KL loss $L_{KL}$:

\begin{equation}
    L_s = \lambda_{KL} L_{KL} + L_R
\end{equation}

The KL loss $L_{KL}$ is the Kullback-Leibler divergence between the encoded $z$ from the encoder and $N(0, 1)^{128}$. The reconstruction loss $L_R$ is the negative log-likelihood of the ground-truth strokes under the GMM and a categorical distribution parametrized by the model. We refer interested readers to a detailed description of $L_{s}$ in the original Sketch-RNN paper~\citep{sketchrnn}.
The initial learning rate of the training procedure was $lr = 1.0 \times 10^{-3}$ and exponentially decayed to $1.0 \times 10^{-5}$ at a rate of $0.9999$. $\lambda_{KL}$ was initially $0.01$ and exponentially increased to $0.5$\footnote{For some object categories, we found that increasing the KL weight to 1.0 improves the authors' perceived quality of generated sketches.} at a rate of $0.99995$. The models were also trained with gradient clipping of $1.0$.

\subsection{Results}
To compare the effectiveness of \systemname at generating scene sketches with existing models and human-level performance, we quantitatively evaluated its performance in an iterative scene authoring task. Moreover, as \systemname uses generative models to produce object sketches, we qualitatively evaluated a large number of examples generated by the two components of \systemnamenospace.

\subsubsection{Composition Modification State-of-the-art}
To evaluate the output of the Composition Proposer against the models introduced with the CoDraw dataset, we adapted its output to match that expected by the well-defined evaluation metrics \crd{proposed by} the original CoDraw paper~\citep{codraw}.
The original task described in the CoDraw paper involves only proposing and modifying high-level object representations in scenes agnostic to their appearance.
The performance of a ``Drawer'' (a human or machine which generates scene compositions) can be quantified by a similarity metric \sig{constrained between 0 and 5} (higher is more similar) \crd{by comparing properties of and relations between objects in the generated scene and objects in the ground truth from the dataset.}

Running our Composition Proposer on the CoDraw test set, we achieved an average similarity metric of $3.55$. This exceeded existing state-of-the-art performance (Table~\ref{tab:sota}) on the iterative scene authoring task using replayed text instructions (script) from CoDraw.

\begin{table}[h]
    \centering
  \caption{Test Set Performance of Various Models on the CoDraw Task}
  \label{tab:sota}
  \begin{tabular}{ccc}
    \toprule
    Teller & Drawer & Similarity $\uparrow{}$ (out of 5)\\
    \midrule
    Script & \textbf{\systemname{}} & \textbf{3.55}\\
    Script &  Neural Network~\citep{codraw} & 3.39\\
    Script & Nearest-Neighbour~\citep{codraw} & 0.94 \\
    \midrule
    Script & Human & \textbf{3.83}\\
  \bottomrule
\end{tabular}
\end{table}

\begin{figure}[h]
  \centering
  \includegraphics[width=0.9\linewidth]{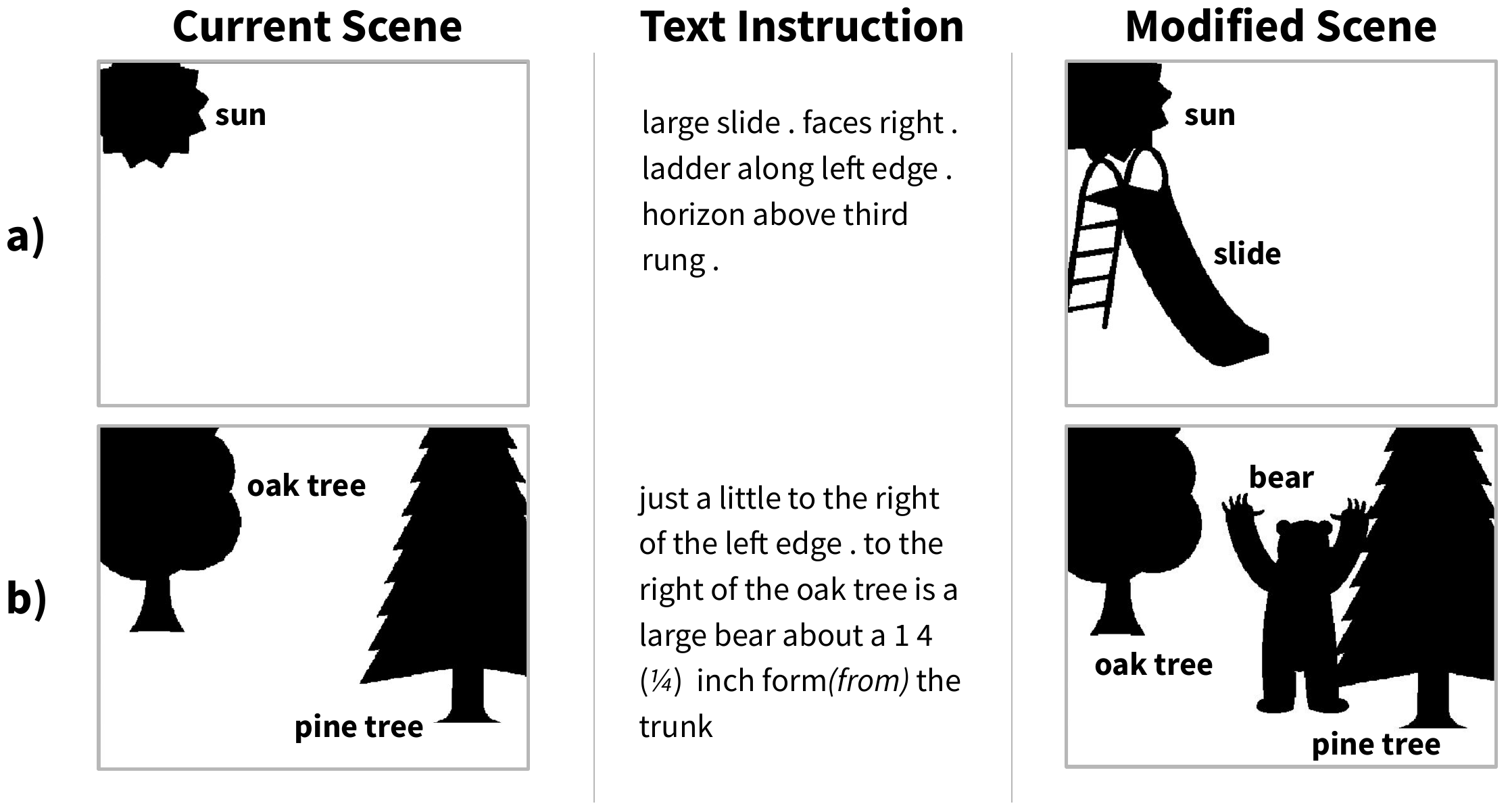}
  \caption{Example Scenes for the Scene Layout Modification Task. The Composition Proposer was able to improve state-of-the-art performance for modifying object representations in scene compositions.}
  \label{fig:composition}
\end{figure}

To provide an illustrative example of our Composition Proposer's output on this task, we visualize two example scenes generated from the CoDraw validation set in Figure~\ref{fig:composition}. 
In \crd{scene a)}, the Composition Proposer extracted the class (slide), direction (faces right), and position relative to parts of the object (ladder along left edge) \crd{from the text instruction}, to place a slide in the scene.
Similarly, it was able to place the bear in between the oak and pine trees \crd{in scene b)}, with the bear touching the left edge of the pine tree.
It is important to note the Composition Proposer completely regenerates the entire scene at each conversation turn. This means it correctly preserved object attributes from previous scenes while making the requested modifications from the current turn. In these instances, the sun in \crd{scene a)} and the trees in \crd{scene b)} were left mostly unchanged while other attributes \crd{of} the scenes were modified.

\subsubsection{Sketches with \crd{Clip Art Objects} as Mask and Ratio Guidance}
The Object Generators are designed to generate sketches which respect high-level scene \crd{layout} information under the guidance of the mask and \crd{aspect} ratio conditions.
To inform generated object sketches \crd{with pose suggestions from scene composition layouts}, we built outline masks from clip art \crd{objects} and computed aspect ratios using the same method as building them for training sketches described in Section \ref{sec:object-sketchers}.
We demonstrate the Object Generators' performance in two important scenarios that allow \systemname to adapt to specific subclass and pose contexts.

\begin{figure}[h]
  \centering
  \includegraphics[width=0.9\linewidth]{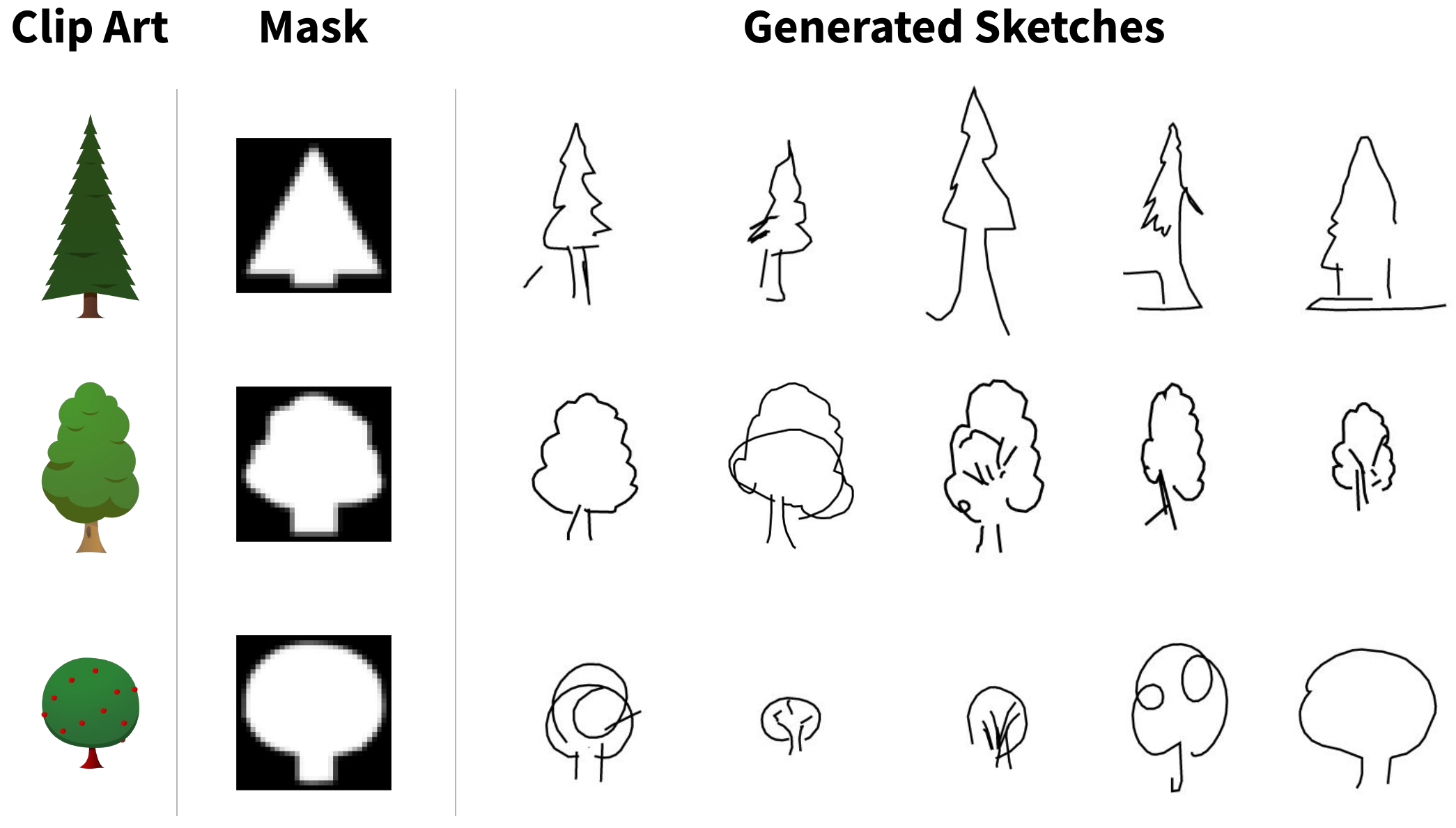}
  \caption{Sketch Generation Results \crd{of Trees} Conditioned on Masks. The Object Generator was able to sketch trees \crd{of} three different classes based on mask and aspect ratio inputs.}
  \label{fig:sketch-tree}
\end{figure}

\paragraph{\emph{Generating objects for closely related classes}}
While the Composition Proposer classifies objects as one distinct class out of 58, some of these classes are closely related and are not differentiated by the Object Generators. In these cases, object masks can be used by \crd{an} Object Generator to effectively disambiguate the desired output subclass.
For instance, the Composition Proposer generates trees as one of three classes: Oak tree (tall and with curly edges), Apple tree (round and short), and Pine tree (tall and pointy); while there is only a single Object Generator trained on a general class of \crd{all types of} tree objects. We generated three different masks and aspect ratios based on three clip art images and used them as inputs to a single tree-based Object Generator to generate appropriate tree objects (by sampling $z \sim N(0, 1)^{128}$).
The Object Generator was able to sketch trees with configurations corresponding to input masks from clip art \crd{objects} (Figure~\ref{fig:sketch-tree}). The generated sketches for pine trees were pointy; for apple trees, had round leaves; and for oak trees, had curvy edges.

\begin{figure}[h]
  \centering
  \includegraphics[width=0.9\linewidth]{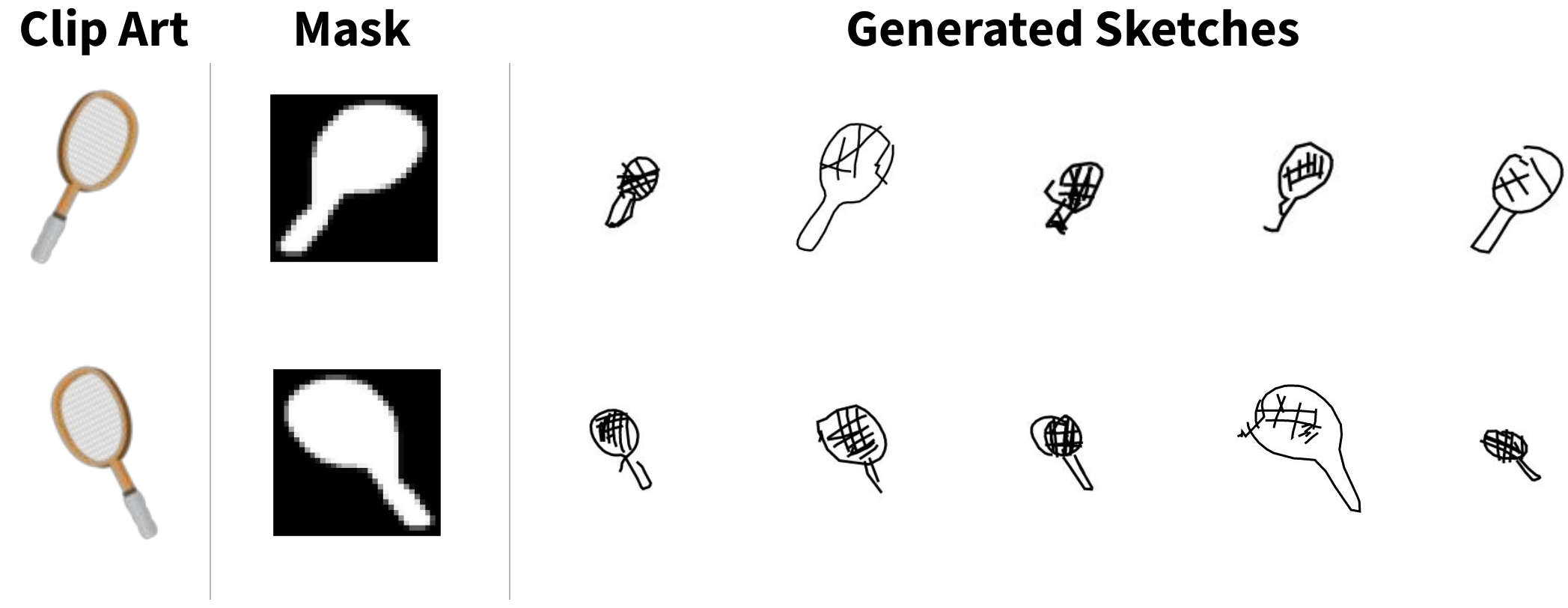}
  \caption{Sketch Generation Results \crd{of Racquets} Conditioned on Masks. The Object Generator was able to sketch racquets at two orientations consistent to the masks.}
  \label{fig:sketch-racquet}
\end{figure}

\paragraph{\emph{Generating objects with direction-specific poses}}
The Composition Proposer can specify the horizontal orientation of the objects (pointing left or right). As such, the Object Generators are required to sketch horizontally asymmetric objects (e.g., racquets, airplanes) with specific poses to follow user\crd{s'} instructions. We show the ability of an Object Generator to produce racquets at various orientations in Figure \ref{fig:sketch-racquet}.
The generated racquet sketches conformed to the orientation of the mask, facing the specified direction at similar angles.

\label{sec:complete}
\subsubsection{Complete Sessions with Composition Layouts and Sketches}
\begin{figure}[htp]
  \centering
  \includegraphics[height=0.93\textheight, keepaspectratio]{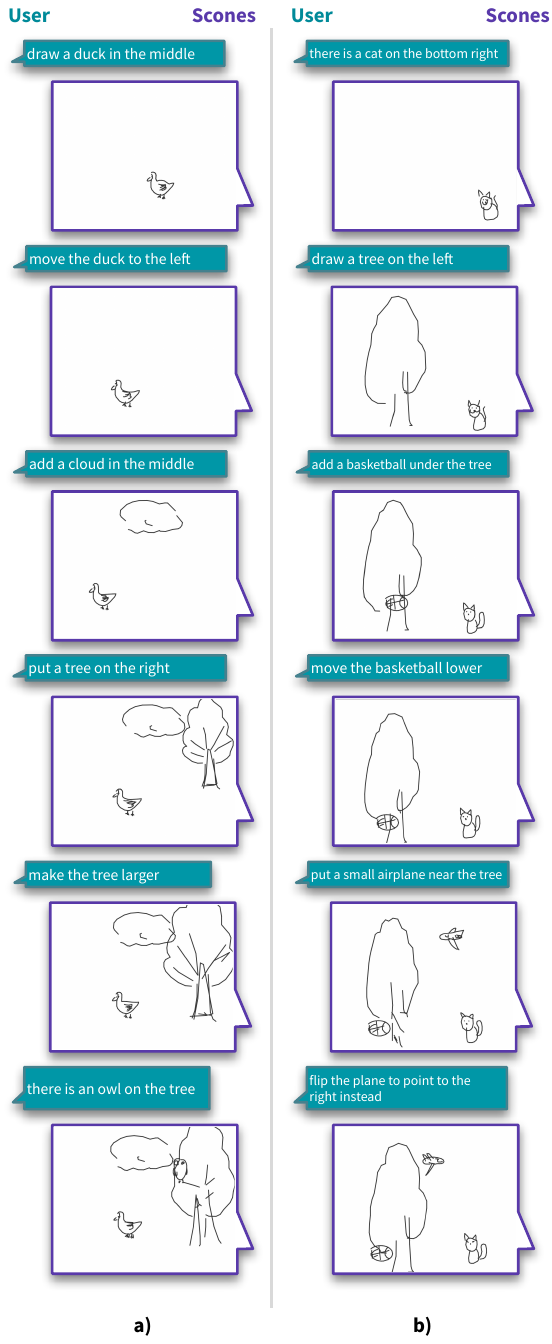}
  \caption{Complete Sketching Sessions with \systemname \crd{curated by the authors.}
  }
  
  \label{fig:complete}
\end{figure}

We show the usage of \systemname in six turns of conversation from multiple sessions in Figure~\ref{fig:complete}. \crd{We curated these sessions by interacting with the system ourselves to demonstrate various capabilities of \systemnamenospace.} In \crd{session a)}, \systemname was able to draw and move the duck to the left, sketch a cloud in the middle, and place and enlarge the tree on the right, following instructions issued by the user. In \crd{session b)}, \systemname was similarly able to place and move \crd{a cat, a tree, a basketball and an airplane}, but at different positions from \crd{session a)}.
For instance, the tree was placed on the left as opposed to the right, and the basketball was moved to the bottom. We also show the ability of \systemname to flip objects horizontally in \crd{session b)}, such that the plane was flipped horizontally and regenerated given the instructions of ``flip the plane to point to the right instead''. This flipping action demonstrates the Object Generator's ability to generate \crd{objects with the require poses} by only sharing the latent vectors $z$, such that the flipped airplane exhibits similar characteristics as the original airplane. In both sessions, \systemname was able to correlate multiple scene objects, such as placing the owl on the tree in session (a), and basketball under the tree in \crd{session b)}.


\subsubsection{Interpreting Transformer's Attention Maps}
\label{sec:attn}
We can further verify the relationships between text and object representations learned by the model by visualizing attention weights computed by the Transformer model of the Composition Proposer. These weights \crd{also} create the unique possibility of generalizing and prompting for sketches of new objects specified by users.

\begin{figure}[h]
  \centering
  \includegraphics[width=\linewidth]{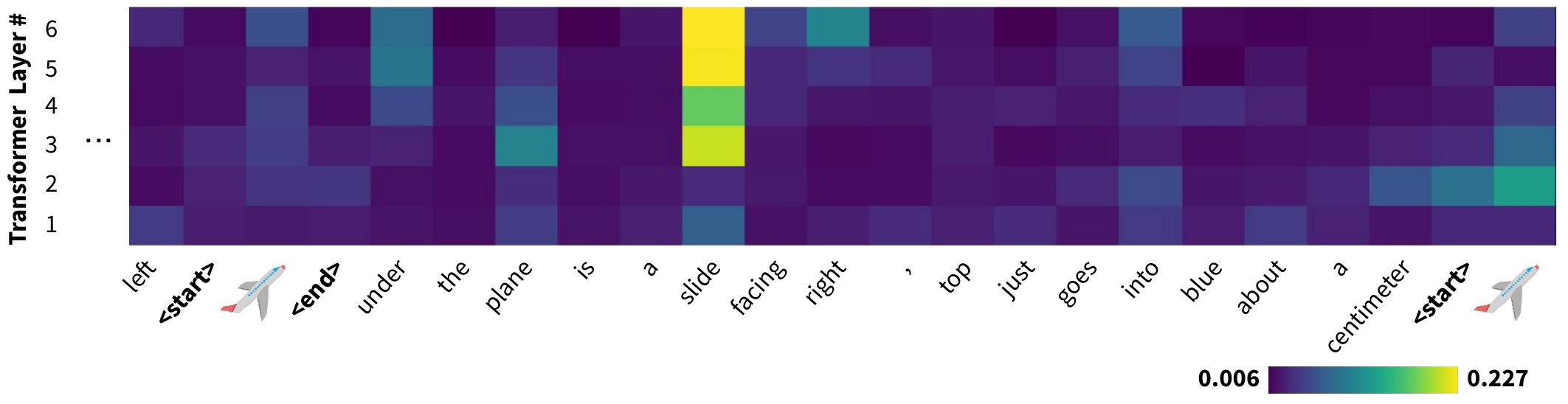}
  \caption{Attention Map of the Transformer across Object and Text Tokens for the Generation of Slide in the Second Turn of Conversation. We observed that the Transformer model attended to the corresponding words and objects \crd{related to the newly generated `slide' object.}}
  \label{fig:attn-second}
\end{figure}

The Transformer model in the Composition Proposer uses masked self-attention to attend to scene objects and instructions from previous time steps most relevant to generating the object specification at the current time step.
We explore the attention weights of the first two turns of a conversation from the CoDraw validation set.
In the first turn, the user instructed the system, ``top left is an airplane medium size pointing left''.
When the model generated the first object, it attended \crd{to} the ``airplane'' and ``medium'' text tokens to select class and output size.
In the second turn, the user instructed the model to place a slide facing right under the airplane.
The model similarly attended to the ``slide'' token the most, it also \crd{significantly} attended to the ``under'' and ``plane'' text tokens\crd{, and the airplane object}. These objects and tokens are important for situating the slide object at the desired location relative to the existing airplane object (Figure~\ref{fig:attn-second}).

\begin{figure}[h]
  \centering
  \includegraphics[width=0.7\linewidth]{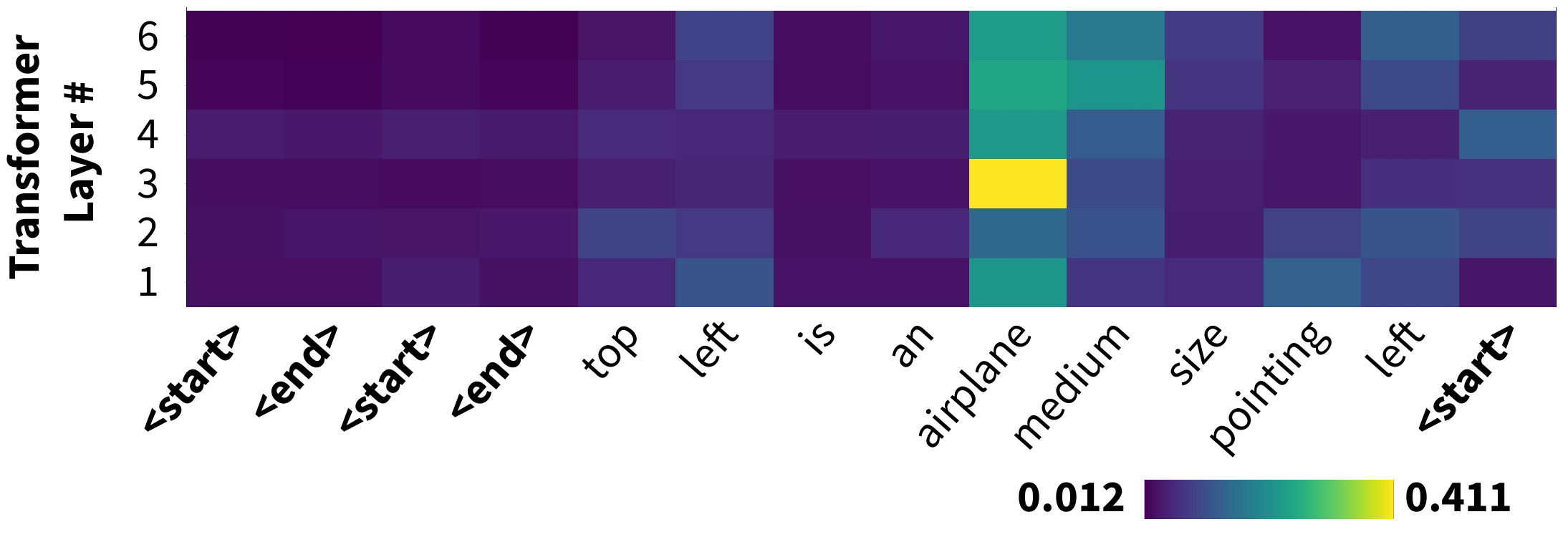}
  \caption{Attention Map of the Transformer across Object and Text Tokens for the Generation of an Airplane, the First Object in the Scene.}
  \label{fig:attn-first}
\end{figure}

\begin{figure}[h]
  \centering
  \includegraphics[width=\linewidth]{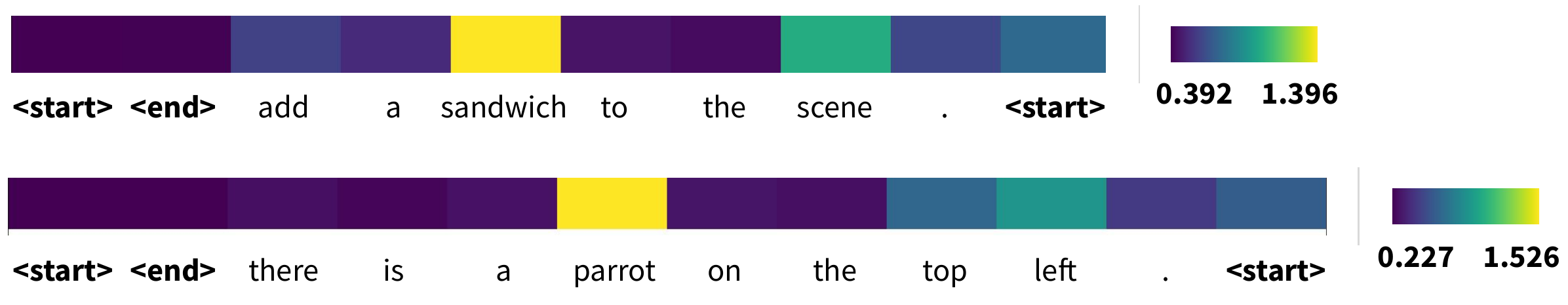}
  \caption{Attention Map of the Transformer for Text
  \crd{Instructions} that Specify Unseen Objects.}
  \label{fig:attn-ooc}
\end{figure}

These attention weights could potentially be used to handle unknown scene objects encountered in instructions.
When the model does not output any scene objects, but only a $o_e$ (scene end) token, we can inspect the attention weights for generating this token to identify a potentially unknown object class, and ask the user for clarification. For example, when a user requests an unsupported class, such as a `sandwich' or  `parrot' (Figure \ref{fig:attn-ooc}), \crd{\systemname could identify this unknown object by taking the text token with the highest attention weight, and prompting the user to sketch it by name.}

\subsection{Exploratory User Evaluation}
\label{sec:user-study}
To determine how effectively \systemname can assist users in creating sketches from natural language, we conducted an exploratory evaluation of \systemnamenospace. We recruited 50 participants from English-speaking countries on Amazon Mechanical Turk (AMT) for our study. We collected quantitative and qualitative results from user trials with \systemnamenospace, as well as suggestions for improving \systemnamenospace. Participants were given a maximum of 20 minutes to complete the study and were compensated \$3.00 USD. Participants were only allowed to complete the task once.

\subsubsection{Method}
\crd{The participants were asked to recreate one of five randomly chosen target scene sketches by providing text instructions to \systemname in the chat window.}
Each target scene had between four and five target objects from a set of 17 possible scene objects. Participants were informed that the final result did not have to be pixel perfect to the target scene, and to mark the sketch as complete once they were happy with the result. Instructions supplied in the chat window were limited to 500 characters, and submitting \crd{an instruction} was considered as taking a ``turn''. The participants were only given the sketch strokes of the target scene without class labels, to elicit natural instructions. 

\begin{figure}[h]
  \centering
  \includegraphics[width=\linewidth]{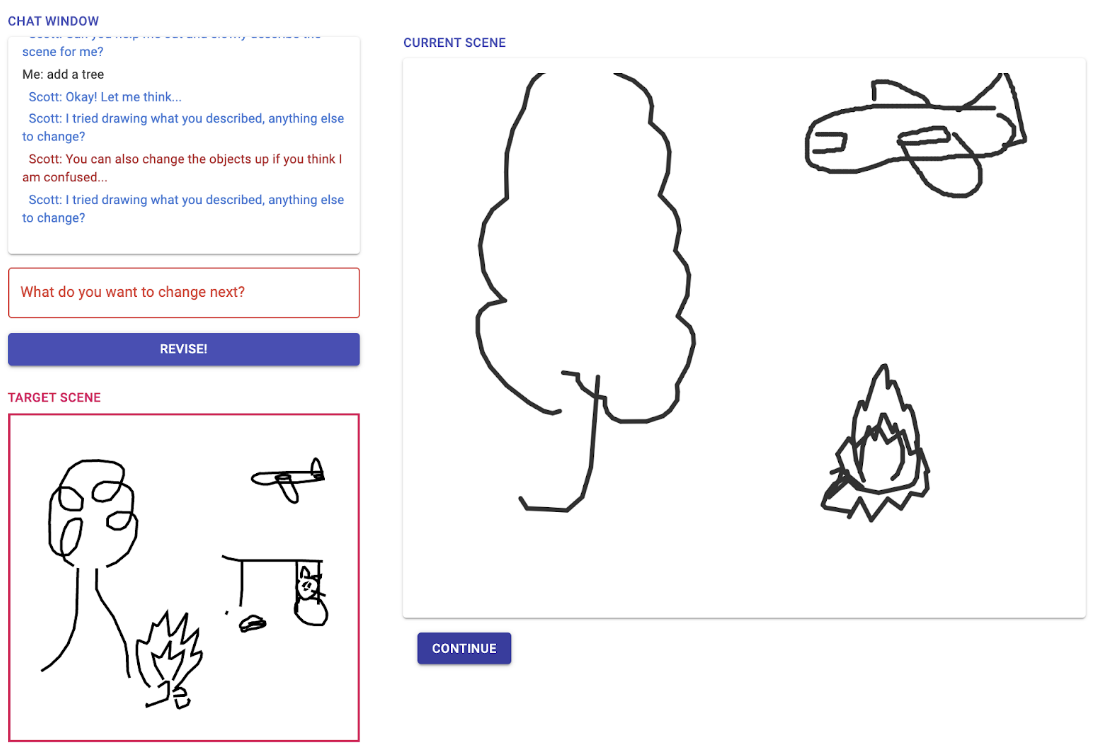}
  \caption{Screenshot of \systemnamenospace's Evaluation User Interface.}
  \label{fig:study-interface}
\end{figure}

Participants were first shown a short tutorial describing the canvas, chat interface, and target scene in the \systemname interface (Figure~\ref{fig:study-interface}), and were asked to give simple instructions in the chat window to recreate the target scene.
\crd{Only two sample instructions were given in the tutorial: ``add a tree'', and ``add a cat next to the table''.} 
At each turn, participants were given the option to redraw objects which remained in the scene for over three turns using a paintbrush\crd{-based} interface. After completing the sketch, participants filled out an exit survey with likert-scale questions on their satisfaction at the sketch and enjoyment of the system, \crd{and open-ended feedback on the system.} 

\subsubsection{Results}

\paragraph{\emph{Participants Satisfied with Sketches, Enjoyment Was Bimodal}}
Participants were generally satisfied with their final sketches ($\mu = 3.38, \, \sigma = 1.18$), and enjoyed the task ($\mu = 4.0, \, \sigma = 1.12$).
In open-ended feedback, participants praised \systemnamenospace's ability to parse their instructions: \textit{``it was able to similarly recreate the image with commands that I typed'' (P25); ``I liked that it would draw what I said. it was simple and fun to use'' (P40).}
Some participants even felt \systemname was able to \emph{intuitively} understand their instructions. P15 remarked, \textit{``I thought it was cool how quickly and intuitively it responded,''} while P35 said, \textit{``It had an intuitive sense of what to draw, and I did not feel constrained in the language I used''.}

\begin{figure}[h]
  \centering
  \includegraphics[width=0.8\linewidth]{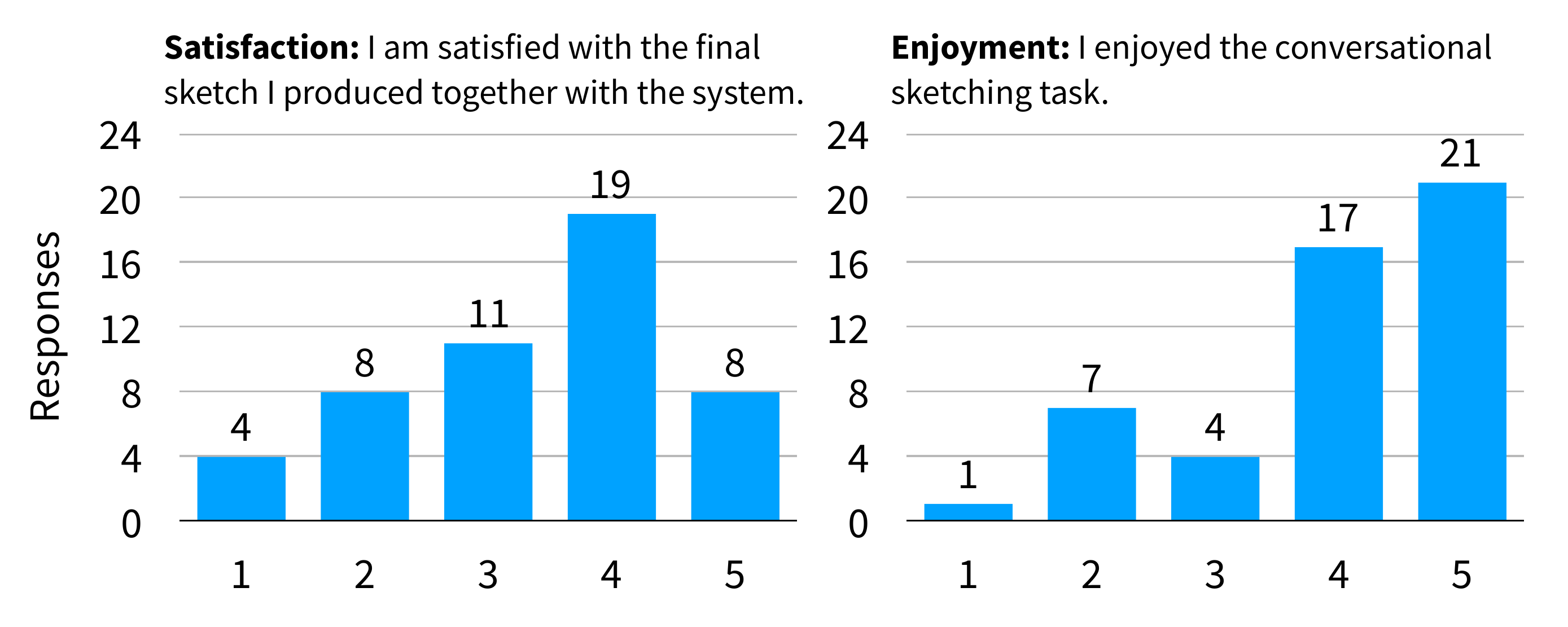}
  \caption{Survey Results from User Sessions with Scones.}
  \label{fig:likerts}
\end{figure}

While enjoyment was high on average, we found responses to enjoyment followed a bimodal distribution (Figure~\ref{fig:likerts}). By reviewing qualitative feedback and instructions to \systemnamenospace, we observe that many instances of low enjoyment (score $\leq 2$) come from class confusion in target scene sketches. Some participants confused the tent in a target scene as a ``pyramid'' in their instructions, which \systemname does not support: \textit{``There is a pyramid on the left side a little ways up from the bottom'' (P44).} P49 tried five times to add a ``pyramid'' to the scene.

P17, who strongly disagreed with enjoying the task (1/5), faced repeated class confusion issues, mentioning, \textit{``it was very frustrating that it wouldn't draw the circle by the cloud \dots It wouldn't draw anything besides the plane, cloud, tent, and fire. Was that not a person up by the cloud?''} \systemname does not support ``circle'' or ``person'' classes---the target sketch had the sun next to the cloud. When \systemname is asked to draw an unsupported object, the canvas will be left unchanged. Providing participants \crd{with} an explicit list of classes in the target image or adding error messages could mitigate these frustrations. Furthermore, attention-based methods mentioned in Section \ref{sec:attn} could be used when an unrecognized class is detected to prompt users to provide sketch strokes with corresponding labels.

\begin{figure}[h]
  \centering
  \includegraphics[width=0.7\linewidth]{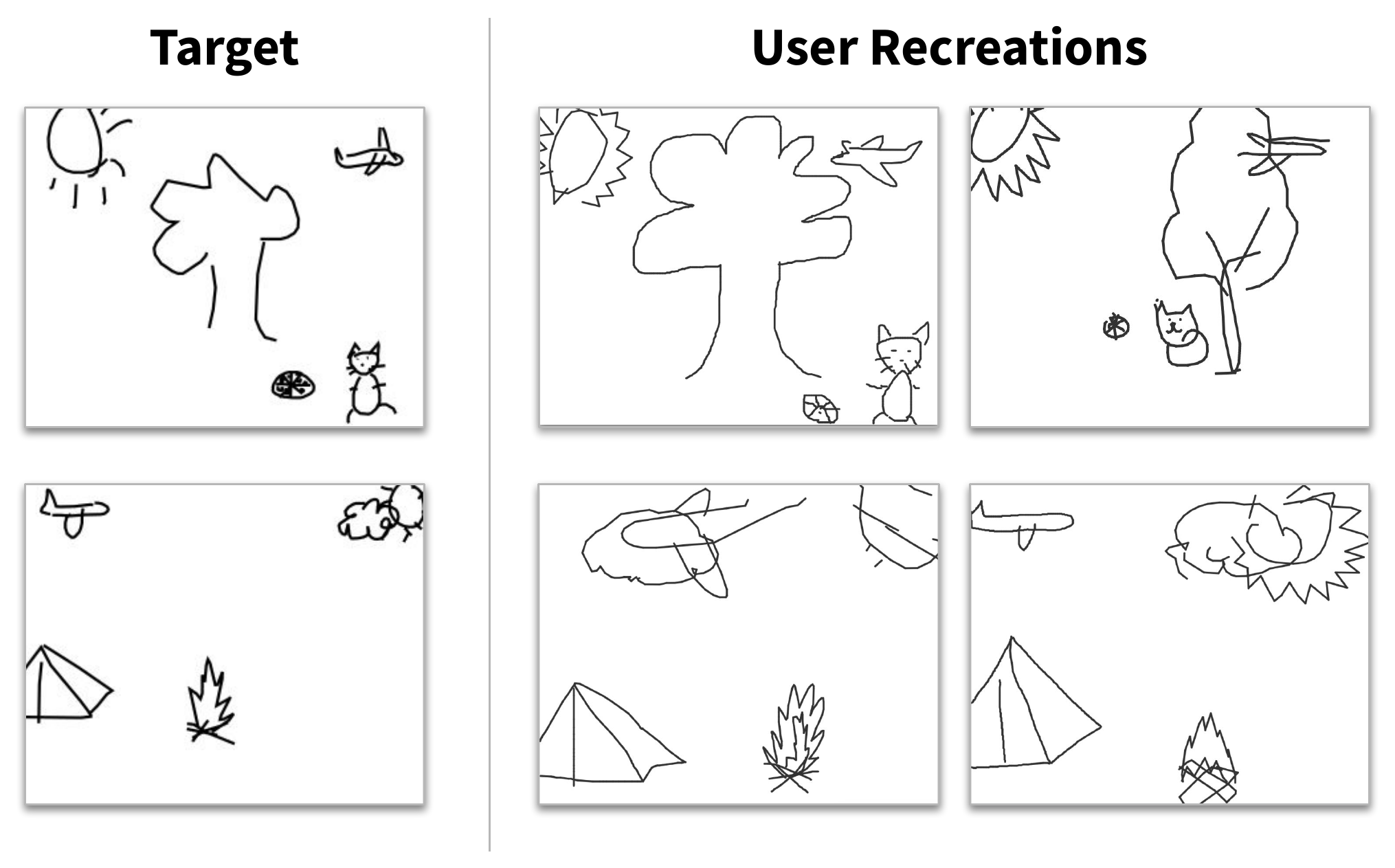}
  \caption{Recreated Scenes during the User Study. Users combined \systemnamenospace-generated outputs with their own sketch strokes to reproduce the target scenes presented to them.}
  \label{fig:recreations}
\end{figure}

\paragraph{\emph{Participants Communicate with Scones at Varying Concept Abstraction Levels}}
On average, participants completed the sketching task in under 8 turns ($\mu = 7.56, \, \sigma = 3.42$), with a varied number of tokens (words \crd{in instructions}) per turn ($\mu = 7.66, \, \sigma = 3.35$). Several participants only asked for the objects themselves (turns delimited by commas): \textit{``helicopter, cloud, swing, add basketball'' (P25)}.
Other participants made highly detailed requests: \textit{``There is a sun in the top left, There is an airplane flying to the right in the top right corner, There is a cat standing on it's hind legs in the bottom right corner, Move the cat a little to the right, please, \dots'' (P14).}
Participants who gave instructions at the expected high-level detail produced satisfying results, \textit{``draw a tree in the middle, Draw a sun in the top left corner, A plane in the top right, A cat with a pizza under the tree'' (P32).}  The recreation of this participant is shown on the top right of Figure~\ref{fig:recreations}.

The longest conversations were often from participants with mismatched expectations for \systemnamenospace, who repeated commands: \textit{``Draw a cloud in the upper left corner with three round edges., Change the cloud to have 3 round edges., Draw only 3 round waves around the edge of the cloud., \dots Draw a snowman to the left of the table., \dots Draw a circle touching the middle circle., \dots'' (P23).} This trial reflects the need for \systemname to make clearer expectations of input to users. \crd{P23's 16-instruction session contains} expectations for the system to modify low-level aspects of the sketches (changing the number of edges in the cloud), exhibits class confusion (snowman and circles with shovel), and has mismatched concept abstraction levels (drawing a shovel versus constructing a shovel from visual primitives, i.e., circles). A potentially simple mitigation for these hurdles would be to introduce more detailed tutorial content for a wider deployment of \systemnamenospace.

\paragraph{\emph{\systemname as a Tool for Collecting Iterative Sketching Data}}
The results of our study show significant potential for \systemname to be used as a Game With a Purpose (GWAP)~\citep{gwap} to collect sketch critiques (natural language specified modifications to an input \crd{sketch} to match a target sketch) and user-generated sketch strokes. 26 ($52\%$ of) participants redrew objects in their sketches when prompted ($\mu = 0.98, \, \sigma = 1.19$), and participants who redrew objects expressed their appreciation for this feature: \textit{``I liked that I could redraw the image'' (P48); ``I liked being able to draw parts myself because it was relaxing and I felt I was more accurate'' (P11).} Most participants who redrew objects also kept output from \systemname in their final sketches, reflecting \systemnamenospace's potential as a mixed-initiative design tool. Redrawing was voluntary in our task, and these results suggest \systemname may be useful for collecting user-generated sketches in addition to natural language critique in a GWAP. Further motivating this application, 14 participants described the task as ``fun'' in open-ended feedback, e.g., \textit{``This was a very fun task'' (P23); ``[I liked] Playing the game and describing the drawing. It was fun!'' (P42).}

\subsubsection{Participants' Feedback for Improving \systemnamenospace}
Participants offered suggestions for how they would improve \systemnamenospace, providing avenues for future work.

\paragraph{\emph{\crd{Object} Translations and Spatial Relationships}}
A major theme of dissatisfaction came from the limited ability of our system to respond to spatial relationships and \crd{translation-related} instructions at times: 
\textit{``It does not appear to understand spatial relationships that well'' (P35); ``you are not able to use directional commands very easily'' (P11).}
These situations largely originate from the CoDraw dataset~\citep{codraw}, in which users had a restricted view of the canvas, resulting in limited relative spatial instructions. This limitation is discussed further in Section \ref{sec:future_data}.

To improve the usability of \systemnamenospace, participants suggest its interface could benefit from \crd{the addition of} direct manipulation features, such as selecting and manually transforming objects in the scene: \textit{``I think that I would maybe change how different items are selected in order to change of modify an object in the picture. (P33); ``ma\crd{yb}e there should be a move function, where we keep the drawing the same but move it'' (P40).}
Moreover, some participants also recommended adding an undo feature, \textit{``Maybe a separate button to get back'' (P31)}, or the ability to manually invoke \systemname to redraw an object, \textit{``I'd like a way to ask the computer to redraw a specific object'' (P3)}. These features could help participants express corrective feedback to \systemnamenospace, potentially creating sketches that better match their intent.

\paragraph{\emph{More Communicative Output}}
Some participants expected \systemname to provide natural language output and feedback to their instructions.
Some participants asked questions directly to elicit \systemnamenospace's capabilities: \textit{``In the foreground is a table, with a salad bowl and a jug of what may be lemonade. In the upper-left is a roughly-sketched sun. Drifting down from the top-center is a box, tethered to a parachute., Did you need me to feed you smaller sentences? \dots'' (P38)}.
P23 explicitly suggested users should be able to ask \systemname questions to refine their intentions: \textit{``I would like the system to ask more questions if it does not understand or if I asked for several revisions. I feel that could help narrow down what I am asking to be drawn''.}
Other participants used praise between their sketching instructions, which could be used as a cue to preserve the sketch output and guide further iteration: \textit{``\dots Draw an airplane, Good try, Draw a table \dots'' (P1); ``Draw a sun in the upper left corner, The sun looks good! Can you draw a hot air balloon in the middle of the page, near the top? \dots'' (P15).}
Providing additional natural language output and prompts from \systemname could enable users to refine \systemnamenospace's understanding of their intent and understand the system's capabilities. A truly \emph{conversational} interface with a sketching support tool could pave the way for advanced mixed-initiative collaborative design tools.

\section{Limitations and Future Research Opportunities}
\label{sec:discussion}
Through the development of a sketch dataset and two sketch-based creativity support systems, we identified some limitations of our work to date and opportunities for future work in this area.

\subsection{Dataset Scale and Match}
\label{sec:future_data}
\begin{figure}
\centering
  \includegraphics[width=0.5\columnwidth]{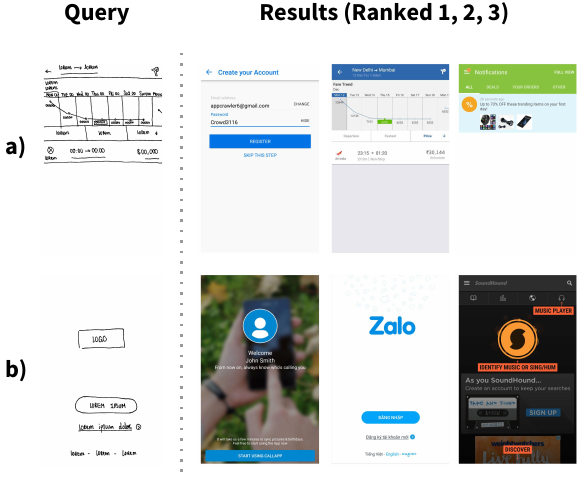}
  \caption{Failure Modes of UI Retrieval using Swire. Swire failed to understand a) custom and b) colorful UI elements.}~\label{fig:query_fail}
\end{figure}

One crucial area for improvement for sketch-based application is the scale and quality of the datasets. Although the dataset we described in Section \ref{sec:swire_data} can be used to train a sketch-based UI retrieval model described in Section \ref{sec:swire_sys}, we observed that the performance of the system has been limited by the diversity and complexity of sketches and UIs in the dataset. This is demonstrated by two major modes of failure in Swire when it struggles to handle rare, custom UI elements as exhibited by Example a in Figure \ref{fig:query_fail}, and fails to understand UIs with diverse colors, such as those with image backgrounds in Example b in Figure \ref{fig:query_fail}. We believe with increased amount of training data, Swire can better generalize to more complex and colorful UIs.

Similarly, we observe that Scones (Section \ref{sec:scones_sys}) has been constrained by the differences between the task protocol used to collect \crd{the} CoDraw \crd{dataset} (the dataset Scones was trained on) and the user interactions in \systemnamenospace. Each conversation in CoDraw only offers the Teller one chance to `peek' at the Drawer's canvas, which significantly decreases the number of modifications made to existing scene objects. As a result, \systemname performs well at adding \crd{objects of} correct classes at appropriate locations and sizes, but is not as advanced at modifying or removing objects. Moreover, the current dataset isn't end-to-end such that there is only a small, fixed number of styles of sketched objects, which reduces Scones' ability in handling stylistic instructions. The ideal data to train this system on shall directly record iterative master-apprentice interactions in creating, critiquing, modifying and removing highly variable sketched scene objects at a large scale. Nevertheless, these datasets are considered to be difficult to collect due to the high sketching skill requirement for crowdworkers~\citep{sketchyscenes}, such that even single-turn multi-object sketched scenes are difficult for crowdworkers to create. 

We believe a significant research direction is to enable the collection of legitimate sketching data at a large-scale, especially in domains that require prior expertise such as UI and mechanical design. This can be achieved by either lowering the skill barrier, or by using tools like Scones and Swire to support parts of joint tasks. For example in the case of Scones, we can lower the skill barrier by decomposing each scene into \crd{object components}, allowing crowdworkers to only sketch a single object at a time in context. Alternatively, future research could explore other means of data collection beyond the crowd. Tools that are used for joint tasks provide natural incentives for designers during realistic design use-cases, and can allow live collection of realistic design data from professional users. Researchers can also investigate sourcing sketches from students of sketching courses at various institutions offering design education who would possess higher sketching expertise.

\subsection{Integration with Applications in Real Usage Scenarios}
Another significant area of research is to create systems that tightly integrate with design and artistic applications in realistic use-cases. While many current research projects (including those described in this book chapter) demonstrate deep-neural-networks' capability and potential in supporting design applications, these applications are currently rough prototypes that are not yet suitable for everyday use by designers and artists. Further developing these applications and exploring how they integrate into design and artistic processes will reveal important usability issues and inform future design and implementation choices of similar tools. For instance, to successfully support UI design processes with Swire, we need to carefully consider the visual representation of UI examples in the application and the underlying datasets to be queried. 

Moreover, some of the capability of these tools can be best demonstrated when applied to professional domains. For instance, \systemname could participate in the UI/UX design process by iteratively suggesting multiple possible modifications of UI design sketches according to design critique. To enable this interaction, we could consider complete UI sketches as `scenes' and UI components as `scene objects'. \systemname could be trained on this data along with text critiques of UI designs to iteratively generate and modify UI mockups from text. To allow \systemname to generate multiple design candidates, we can modify the current architecture to model probabilistic outputs for both sketch strokes (which is currently probabilistic) and scene object locations. While datasets of UI layouts and components, such as those presented in Section \ref{sec:swire_data}, suggest this as a near possibility, this approach may generalize to other domains as well, such as industrial design. Nevertheless, this requires significant data support by solving the issues mentioned in Section \ref{sec:future_data}.


\section{Conclusion}
This chapter presented three key aspects of developing deep-learning-driven, sketch-based creativity support tools. First, we collect the first large-scale dataset of sketches corresponding to UIs. Second, we develop a sketch-based UI retrieval technique that enables designers to interact with large-scale UI datasets using sketches. Third, we built a deep-learning-driven system that supports the novel interaction of generating scenes of sketched objects from text instructions. The dataset we collected supported our deep-learning-based tools, and we showed qualitatively and quantitatively 
that our systems can support targeted design and artistic applications. We further outlined several areas of future research opportunities and hope the documentation of our development experience and the release of our dataset can spur future research in this area.

Our ultimate goal of pursuing this line of research is to provide users of all levels of sketching expertise with relevant materials and computational resources to focus on creative and innovative tasks in creative processes. We also hope these projects can provide entirely new means for creative expression and rapid ideation. We are excited to continue designing for this future of design, art, and engineering.



\begingroup
\bibliography{bibliography}
\endgroup
\end{document}